\documentclass[a4paper, 12 pt]{article}
\usepackage[margin = 0.75 in, includefoot]{geometry} 
\usepackage{setspace} 
\usepackage{amsmath, amssymb, amsthm, mathrsfs} 
\usepackage[title]{appendix} 
\usepackage[affil-it]{authblk} 
\usepackage{blkarray, bigstrut} 
\usepackage{chngcntr} 
\usepackage{enumerate} 
\usepackage{graphicx} 
\usepackage{multicol} 
\usepackage{multirow} 
\usepackage[numbers]{natbib} 
\usepackage{quotes} 
\usepackage{xcolor} 
\usepackage{hyperref} 

\newcommand\topstrut[1][1.2ex]{\setlength\bigstrutjot{#1}{\bigstrut[t]}} 
\newcommand\botstrut[1][0.9ex]{\setlength\bigstrutjot{#1}{\bigstrut[b]}} 
\theoremstyle{remark}
\newtheorem{remark}{Remark} 
\newcounter{example} 
\newenvironment{example}[1][]{\refstepcounter{example}\par
	\noindent \textbf{Example~\theexample. #1}}{\bigskip}
\allowdisplaybreaks 

\title{\textbf{Comparative Analysis of Kinetic \\ Realizations of Insulin Signaling}}

\author[1]{\textbf{Patrick Vincent N. Lubenia}}
\author[1,2,3]{\textbf{Eduardo R. Mendoza}}
\author[1,2,4\thanks{Corresponding author. \\ \indent \hspace{0.01 cm} Emails: angelyn.lao@dlsu.edu.ph, pnlubenia@upd.edu.ph, eduardo.mendoza@dlsu.edu.ph}]{\textbf{Angelyn R. Lao}}

\affil[1]{Systems and Computational Biology Research Unit, Center for Natural Sciences and Environmental Research, 2401 Taft Avenue, Manila, 0922, Metro Manila, Philippines}
\affil[2]{Department of Mathematics and Statistics, De La Salle University, 2401 Taft Avenue, Manila, 0922, Metro Manila, Philippines}
\affil[3]{Max Planck Institute of Biochemistry, Am Klopferspitz 18, 82152, Martinsried near Munich, Germany}
\affil[4]{Center for Complexity and Emerging Technologies, 2401 Taft Avenue, Manila, 0922, Metro Manila, Philippines}

\date{}

\begin{document}

\maketitle

\begin{abstract}
	Several studies have developed dynamical models to understand the underlying mechanisms of insulin signaling, a signaling cascade that leads to the translocation of glucose, the human body's main source of energy. Fortunately, reaction network analysis allows us to extract properties of dynamical systems without depending on their model parameter values. This study focuses on the comparison of insulin signaling in healthy state (INSMS or INSulin Metabolic Signaling) and in type 2 diabetes (INRES or INsulin RESistance) using reaction network analysis. The analysis uses network decomposition to identify the different subsystems involved in insulin signaling (e.g., insulin receptor binding and recycling, GLUT4 translocation, and ERK signaling pathway, among others). Furthermore, results show that INSMS and INRES are similar with respect to some network, structo-kinetic, and kinetic properties. Their differences, however, provide insights into what happens when insulin resistance occurs. First, the variation in the number of species involved in INSMS and INRES suggests that when irregularities occur in the insulin signaling pathway, other complexes (and, hence, other processes) get involved, characterizing insulin resistance. Second, the loss of concordance exhibited by INRES suggests less restrictive interplay between the species involved in insulin signaling, leading to unusual activities in the signaling cascade. Lastly, GLUT4 losing its absolute concentration robustness in INRES may signify that the transporter has lost its reliability in shuttling glucose to the cell, inhibiting efficient cellular energy production. This study also suggests possible applications of the equilibria parametrization and network decomposition, resulting from the analysis, to potentially establish absolute concentration robustness in a species.

\bigskip

\end{abstract}

\section{Introduction}

Insulin signaling plays a crucial role in the human body's energy metabolism. Insulin reception triggers other processes that lead to the translocation of glucose, the body's main source of energy \citep{YFBS2019}. However, irregularities that decrease insulin reception can lead to insulin resistance in cells. Researchers have studied the role of insulin resistance in the development of many metabolic disorders such as type 2 diabetes, and of other phenomena in the human body including oxidative stress, inflammation, insulin receptor mutation, and mitochondrial dysfunction \citep{NTNLV2020, ONEASZ2018, TGCCLMOF2020}. To this day, the underlying cause of insulin resistance has not been fully understood since the phenomenon is highly complicated and the underlying mechanisms in healthy and insulin-resistant cells have not yet been identified by researchers dealing with the difficult problem.

Mathematical models have helped the scientific community to better understand insulin signaling in healthy cells \citep{HWDLC2014, QC1991, SSQ2002, WQ2000} and diseased cells \citep{BC2015, BNFBECS2013, NRFBCS2014}. However, the complexity of metabolic insulin signaling have also naturally led to complicated dynamical models. The results generated from these models depend on simulations that rely on numerous parameter values that need to be identified from literature or derived from experimental data. Chemical Reaction Network Theory (CRNT) can address this challenge inherent in many dynamical systems. Hence, analysis using CRNT is a useful approach in studying biological systems.

CRNT translates a system of differential equations into a chemical reaction network (CRN) representation or kinetic realization of the system's network of reactions. A CRN is composed of nonempty finite sets of species, complexes, and reactions. A complex is one which appears as a reactant or a product in a reaction. Upon identification of a kinetics (i.e., an assignment of rate functions to the reactions), one can then talk about the chemical kinetic system.

Constructing a particular kinetic realization of a system and identifying its kinetics allow the extraction of formal properties of the system, independent from parameter values of the model. Working with kinetic realizations also allows us to compare different instances of the same system, e.g., insulin signaling in healthy and insulin-resistant cells. One can also decompose the CRN representation into subnetworks to determine subgroups in the system, a task that is difficult to do with differential equations and biochemical maps. Several papers have taken advantage of the usefulness of CRNT in dealing with complex networks \citep{AJLM2017, FOME2023, HLJK2022, KPDDG2012, LML2022, MRBH2015, SHFE2010, SHFE2011, TG2009a, TG2009b, VLMA2019}.

In this study, we use two systems of ordinary differential equations (ODEs): a model of insulin signaling in healthy cells by Sedaghat et al. \cite{SSQ2002} and a model of insulin signaling in type 2 diabetes by Nyman et al. \cite{NRFBCS2014}. Lubenia et al. \cite{LML2022} have constructed a kinetic realization of the Sedaghat et al. mass action model and have already performed a reaction network analysis of the system. Their key findings are as follows: $(i)$ the underlying network is concordant; $(ii)$ the network of three functional modules discussed by Sedaghat and colleagues form an independent decomposition; and $(iii)$ the network has 8 species with absolute concentration robustness (ACR), one of which is the essential glucose transporter GLUT4 which, coupled with adequate glucose supply, enables reliable cellular energy production. We denote the kinetic realization of the insulin signaling in healthy cell by INSMS (INSulin Metabolic Signaling).

A natural next step is to check what happens to insulin signaling when there is insulin resistance. To this end, we use the Nyman et al. model of insulin signaling in type 2 diabetes. We then perform a comparative analysis of the kinetic realizations of the two states of insulin signaling, an analysis which, to our knowledge, is the first of its kind regarding metabolic insulin signaling. Similar to the Sedaghat et al. model, the Nyman et al. model is also a mass action system, making our reaction network analysis parameter-free. We denote the kinetic realization of the insulin signaling in type 2 diabetes by INRES (INsulin RESistance).

In our comparative analysis, we utilize decomposition theory to identify the subsystems involved in insulin signaling. Decomposing the networks into subnetworks allows us to identify in INRES the three functional modules that also appear in INSMS: insulin receptor binding and recycling, postreceptor signaling, and GLUT4 translocation. Furthermore, the decomposition shows further subsystems that are present in INRES but not in INSMS: the S6 and S6K formation from mTORC, ERK signaling pathway, and nuclear transcription. More importantly, decomposition theory helps identify the subnetworks of INRES which correspond to the ``single mechanism'' that explains insulin resistance in type 2 diabetes according to Nyman et al.: an attenuated positive feedback from mTORC to IRS1.

We also look at the essential properties of INSMS and INRES. The properties considered are essential in the sense that they do not depend on the parameters, i.e., rate constants, used in the mass action models they were derived from. A network (or structural) property is one that can be specified in terms of the network components alone (e.g., concordance). Properties which may change under different dynamic equivalences are called structo-kinetic properties (e.g., monostationarity) while those which are invariant under any dynamic equivalence are called (purely) kinetic properties (e.g., ACR). INSMS and INRES are similar with respect to some network, structo-kinetic, and kinetic properties. But more importantly, this study confirms our expectation that kinetic system differences reveal notable biological differences of the same system under differing states, i.e., healthy and insulin-resistant.

First, the (molecular) species involved in insulin signaling (and consequently the functional modules they constitute) differ strongly. This variation in the number of species involved in INSMS and INRES suggests that when irregularities occur in the insulin signaling pathway, other complexes (and, hence, other processes) get involved, characterizing insulin resistance. This and other properties point to a generally higher complexity of signal processing in the insulin-resistant case.

Second, INRES is discordant, i.e., the concordance of INSMS is lost. According to Shinar and Feinberg \cite{SHFE2013}, concordant networks indicate ``architectures that, by their very nature, enforce duller, more restrictive behavior despite what might be great intricacy in the interplay of many species, even independent of values that kinetic parameters might take''. Thus, the loss of concordance exhibited by INRES suggests less restrictive interplay between the species involved in insulin signaling, leading to unusual activities in the signaling cascade.

Finally, there are no ACR species in INRES, marking the loss of ACR in 6 of its common species with INSMS, including the critical GLUT4 which is responsible for transporting glucose into the glycolytic system. The loss of ACR in GLUT4 suggests that, in an insulin-resistant cell, the transporter has lost its reliability in shuttling glucose to the cell, inhibiting efficient cellular energy production. We discuss in Section \ref{sec:eqParam} possible applications of a network's equilibria parametrization and its finest independent decomposition to potentially establish ACR in a species.

The similarities and differences in the features of the two networks suggest that insulin signaling in healthy and insulin-resistant cells follow different pathways leading to the glucose transporter GLUT4 and its translocation of glucose.

This paper is organized as follows: in Section \ref{sec:insms}, we summarize the results of the reaction network analysis of insulin signaling in healthy cells by Lubenia et al. \cite{LML2022}. In Section \ref{sec:inres}, we construct the kinetic realization of the model of insulin signaling in type 2 diabetes by Nyman et al. \cite{NRFBCS2014}. Section \ref{sec:structural} details the comparative analysis of network properties of INSMS and INRES while Section \ref{sec:kinetic} considers the networks' structo-kinetic and kinetic  properties. Finally, we conclude the paper with a summary in Section \ref{sec:summary}. An Appendix with background on CRNT and other pertinent details of the models considered is provided for the reader.

\section{Reaction Network Analysis of Insulin Signaling in \\ Healthy Cells}
\label{sec:insms}

This section revisits the results of the reaction network analysis of Lubenia et al. \cite{LML2022} of a model of insulin signaling in healthy cells. They constructed a chemical reaction network (CRN) with mass action kinetics based on the insulin signaling model by Sedaghat et al. \cite{SSQ2002}. We denote this kinetic realization by INSMS (INSulin Metabolic Signaling). Tables \ref{tab:INSMSnumbers} and \ref{tab:INSMSproperties} give the network numbers and an overview of the properties of INSMS. The reader may refer to Appendix \ref{app:crnt} for a review of some basic concepts in Chemical Reaction Network Theory (CRNT).

\begin{table}[ht]
    \begin{center}
        \caption{\textbf{Network numbers of INSMS} The high values of species, (reactant) complexes, reactions, (reactant) rank, and (reactant) deficiency quantify the high complexity of insulin signaling in healthy cells; the network numbers indicate that INSMS is a closed network with high reactant diversity; the network is also branching, $t$-minimal, not weakly reversible, and its terminal classes contain points and cycles\\}
        \label{tab:INSMSnumbers}
        \begin{tabular}{@{}lc@{}}
            \hline
            \multicolumn{1}{c}{\textbf{Characteristic}} & \textbf{Value} \\
            \hline
            Species & 20 \\
            Complexes & 35 \\
            Reactant complexes & 24 \\
            Reactions & 35 \\
            Linkage classes & 13 \\
            Strong linkage classes & 24 \\
            Terminal strong linkage classes & 13 \\
            Rank & 15 \\
            Reactant rank & 20 \\
            Deficiency & 7 \\
            Reactant deficiency & 4 \\
            \hline
        \end{tabular}
    \end{center}
\end{table}

\begin{table}[ht]
    \begin{center}
        \caption{\textbf{Overview of properties of INSMS} Some of the network properties listed are described in Table \ref{tab:INSMSnumbers}; INSMS has a positive equilibrium for some sets of rate constants (since it is positive dependent), its reactions have a high degree of linear dependence among each other (due to high deficiency), and is nonconservative and concordant; the key structo-kinetic properties of INSMS include its monostationarity (it cannot have multiple positive equilibria for a set of rate constants) and injectivity; its kinetic properties include the nondegeneracy of its equilibria and the existence of 8 ACR species in the system\\}
        \label{tab:INSMSproperties}
        \begin{tabular}{@{}|c|ll|@{}}
            \hline
            \textbf{Property Class} & \multicolumn{2}{c|}{\textbf{INSMS}} \\
            \hline
            \multirow{5}{*}{Network} & Closed & Not (weakly) reversible \\
             & $t$-minimal & Branching \\
             & \multicolumn{2}{l|}{Terminal classes include points and cycles} \\
              & Positive dependent & Deficiency 7 \\
             & Nonconservative & Concordant \\
            \hline
            Structo-Kinetic & Monostationary & Injective \\
            \hline
            Kinetic & Nondegenerate equilibria & 8 ACR species \\
            \hline
        \end{tabular}
    \end{center}
\end{table}

INSMS is clearly a complex network as it involves 20 species, 35 complexes, and 35 reactions. It is a closed network with high reactant diversity. Viewed as a digraph, INSMS is branching, $t$-minimal, not (weakly) reversible, and its terminal classes contain points and cycles. Based on its stoichiometry, the network is positive dependent, of deficiency 7, nonconservative, and concordant. The authors also observed that INSMS has a nontrivial finest independent decomposition composed of 10 subnetworks. The biological significance of these and further properties are discussed in the succeeding chapters.

One of the key properties implied by the concordance of INSMS is its injectivity. The weak momotonicity of the mass action system implies that the network is also monostationary. Furthermore, all positive equilibria of INSMS are nondegenerate. And one of the most important results of Lubenia and colleagues was the discovery that 8 out of 20 species, including the critical GLUT4, in the insulin signaling in healthy cells has absolute concentration robustness (ACR).

\section{Kinetic Realization of the Nyman et al. Model}
\label{sec:inres}

In this section, we discuss the model of insulin signaling in type 2 diabetes by Nyman et al. \cite{NRFBCS2014}. After an overview of their biological findings, we construct a CRN representation of the dynamical system with mass action kinetics and discuss some of the basic properties of this kinetic realization.

\subsection{Summary of Novel Biological Insights from the Model}
\label{subsec:NymanBiology}

Using data based on human adipocytes from healthy and diabetic individuals, Nyman and colleagues identified three hallmarks of insulin resistance in type 2 diabetes:
\begin{enumerate}
    \item Diminished concentration of insulin receptors;
    \item Low concentration of the glucose transporter GLUT4; and
    \item Impaired feedback from mammalian target of rapamycin complex 1 (mTORC1) to insulin receptor substrate 1 (IRS1).
\end{enumerate}

\begin{remark}
    Hallmarks 1 and 2 are reflected in the loss of concentration robustness of the ACR species in INSMS (discussed in Section \ref{sec:acr}).
\end{remark}

The model the authors used is a modification of an existing system of ordinary differential equations (ODEs) representing insulin signaling in type 2 diabetes constructed by Br\"{a}nnmark et al. \cite{BNFBECS2013}. Nyman et al. added the extracellular signal-regulated kinase (ERK) signaling pathway, an important signaling branch in type 2 diabetes. After comparing observed data and generating simulations, the authors determined that, among the three main hallmarks of insulin resistance, the reduced positive feedback from mTORC1 to IRS1 could explain insulin resistance in all parts of the insulin signaling pathway.

\subsection{The Reaction Network of the Nyman et al. Model}
\label{subsec:inres}

The Nyman et al. model is composed of 32 ODEs (see Appendix \ref{app:nyman} for the system of ODEs and the description of the variables). All 44 reactions in their model are modeled using mass action kinetics.

For easy comparison, the species occurring in both Sedaghat et al. and Nyman et al. models are denoted by the same variable. Variables for those occurring solely in Nyman et al.'s model continue their numbering from those in the Sedaghat et al. model.

\begin{figure}[ht]
    \centering
    \includegraphics[width = 0.8\textwidth]{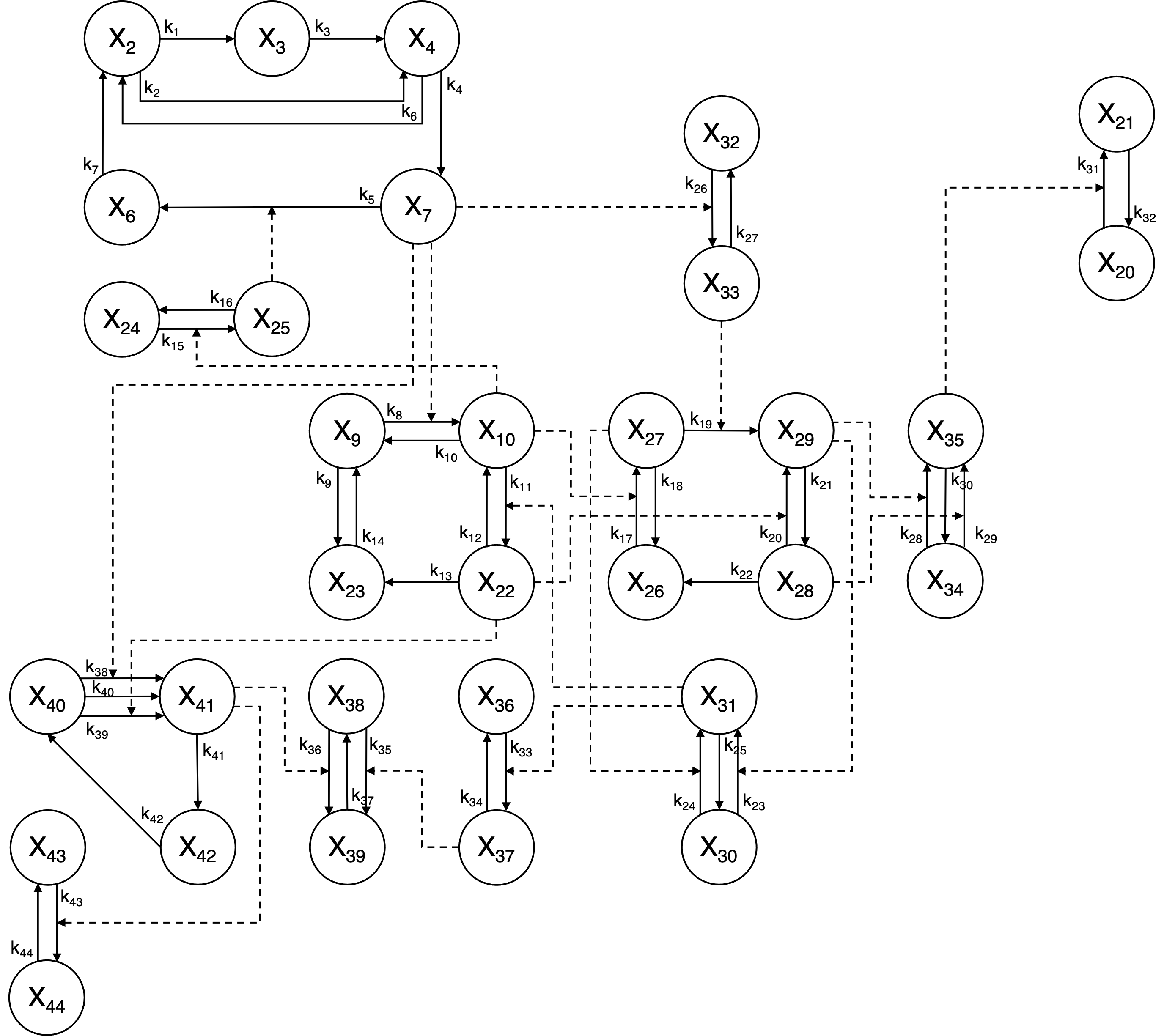}
    \caption{\textbf{Biochemical map of insulin signaling in type 2 diabetes} $X_2$, $X_3$, $X_4$, $X_6$, $X_7$, $X_9$, $X_{10}$, $X_{20}, \ldots, X_{44}$ are the species of the network (see Appendix \ref{app:nyman} for the description of the variables); $k_1, \ldots, k_{44}$ are the rate constants of the reactions; and solid lines represent mass transfer reactions while broken lines represent regulatory reactions; identifying subnetworks in insulin signaling in type 2 diabetes is difficult to do with biochemical maps but can be easily done using kinetic realizations}
    \label{fig1}
\end{figure}

For better visual orientation of the reader, we reconstruct the complete biochemical map of the Nyman et al. mass action system in Figure \ref{fig1}. The supplementary materials of Br\"{a}nnmark et al. \cite{BNFBECS2013} and Nyman et al. \cite{NRFBCS2014} contain details of the construction of the ODEs.

Applying the the Hars-T\'{o}th criterion for mass action system realization, as used in Lubenia et al. \cite{LML2022} and detailed in Chellaboina et al. \cite{CBHB2009}, a kinetic realization of the Nyman et al. model is as follows:

\begin{multicols}{2}
\noindent
\begin{align*}
    & R_{1}: X_2 \rightarrow X_3 \\
    & R_{2}: X_2 \rightarrow X_4 \\
    & R_{3}: X_3 \rightarrow X_4 \\
    & R_{4}: X_4 \rightarrow X_7 \\
    & R_{5}: X_7 + X_{25} \rightarrow X_6 + X_{25} \\
    & R_{6}: X_4 \rightarrow X_2 \\
    & R_{7}: X_6 \rightarrow X_2 \\
    & R_{8}: X_7 + X_9 \rightarrow X_7 + X_{10} \\
    & R_{9}: X_9 \rightarrow X_{23} \\
    & R_{10}: X_{10} \rightarrow X_9 \\
    & R_{11}: X_{10} + X_{31} \rightarrow X_{22} + X_{31} \\
    & R_{12}: X_{22} \rightarrow X_{10} \\
    & R_{13}: X_{22} \rightarrow X_{23} \\
    & R_{14}: X_{23} \rightarrow X_9 \\
    & R_{15}: X_{10} + X_{24} \rightarrow X_{10} + X_{25} \\
    & R_{16}: X_{25} \rightarrow X_{24} \\
    & R_{17}: X_{10} + X_{26} \rightarrow X_{10} + X_{27} \\
    & R_{18}: X_{27} \rightarrow X_{26} \\
    & R_{19}: X_{27} + X_{33} \rightarrow X_{29} + X_{33} \\
    & R_{20}: X_{22} + X_{28} \rightarrow X_{22} + X_{29} \\
    & R_{21}: X_{29} \rightarrow X_{28} \\
    & R_{22}: X_{28} \rightarrow X_{26} \\
    & R_{23}: X_{29} + X_{30} \rightarrow X_{29} + X_{31} \\
    & R_{24}: X_{27} + X_{30} \rightarrow X_{27} + X_{31} \\
    & R_{25}: X_{31} \rightarrow X_{30} \\
    & R_{26}: X_7 + X_{32} \rightarrow X_7 + X_{33} \\
    & R_{27}: X_{33} \rightarrow X_{32} \\
    & R_{28}: X_{29} + X_{34} \rightarrow X_{29} + X_{35} \\
    & R_{29}: X_{28} + X_{34} \rightarrow X_{28} + X_{35} \\
    & R_{30}: X_{35} \rightarrow X_{34} \\
    & R_{31}: X_{35} + X_{20} \rightarrow X_{35} + X_{21} \\
    & R_{32}: X_{21} \rightarrow X_{20} \\
    & R_{33}: X_{31} + X_{36} \rightarrow X_{31} + X_{37} \\
    & R_{34}: X_{37} \rightarrow X_{36} \\
    & R_{35}: X_{37} + X_{38} \rightarrow X_{37} + X_{39} \\
    & R_{36}: X_{38} + X_{41} \rightarrow X_{39} + X_{41} \\
    & R_{37}: X_{39} \rightarrow X_{38} \\
    & R_{38}: X_7 + X_{40} \rightarrow X_7 + X_{41} \\
    & R_{39}: X_{22} + X_{40} \rightarrow X_{22} + X_{41} \\
    & R_{40}: X_{40} \rightarrow X_{41} \\
    & R_{41}: X_{41} \rightarrow X_{42} \\
    & R_{42}: X_{42} \rightarrow X_{40} \\
    & R_{43}: X_{41} + X_{43} \rightarrow X_{41} + X_{44} \\
    & R_{44}: X_{44} \rightarrow X_{43}
\end{align*}
\end{multicols}
This realization coincides with the biochemical map in Figure \ref{fig1}. Moreover, we denote this kinetic realization by INRES (INsulin RESistance) (also $\mathscr{N} = (\mathscr{S}, \mathscr{C}, \mathscr{R})$ with mass action kinetics $K$, set $\mathscr{S}$ of 32 species, set $\mathscr{C}$ of 70 complexes, and set $\mathscr{R}$ of 44 reactions). Tables \ref{tab:INRESnumbers} and \ref{tab:INRESproperties} present the network numbers and an overview of the properties of INRES.

\begin{table}[ht]
    \begin{center}
        \caption{\textbf{Network numbers of INRES} The high values of species, (reactant) complexes, reactions, (reactant) rank, and (reactant) deficiency quantify the high complexity of insulin signaling in insulin-resistant cells; the network numbers indicate that INRES is a closed network ($s < m$) with high reactant diversity ($n_r > s$); the network is also branching ($r > n_r$), $t$-minimal ($t = \ell$), not weakly reversible ($s\ell \neq \ell$), and its terminal classes contain points ($t \neq n - n_r$) and cycles ($n \neq n_r$)\\}
        \label{tab:INRESnumbers}
        \begin{tabular}{@{}lcc@{}}
            \hline
            \multicolumn{1}{c}{\textbf{Characteristic}} & \textbf{Notation} & \textbf{Value} \\
            \hline
            Species & $m$ & 32 \\
            Complexes & $n$ & 70 \\
            Reactant complexes & $n_r$ & 41 \\
            Reactions & $r$ & 44 \\
            Linkage classes & $\ell$ & 31 \\
            Strong linkage classes & $s \ell$ & 65 \\
            Terminal strong linkage classes & $t$ & 31 \\
            Rank & $s$ & 20 \\
            Reactant rank & $q$ & 32 \\
            Deficiency & $\delta$ & 19 \\
            Reactant deficiency & $\delta_p$ & 9 \\
            \hline
        \end{tabular}
    \end{center}
\end{table}

\begin{table}[ht]
    \begin{center}
        \caption{\textbf{Overview of properties of INRES} Some of the network properties listed are described in Table \ref{tab:INRESnumbers}; INRES has a positive equilibrium for some sets of rate constants (since it is positive dependent), its reactions have a high degree of linear dependence among each other (due to high deficiency), and is conservative and discordant; the key structo-kinetic properties of INRES include its monostationarity (it cannot have multiple positive equilibria for a set of rate constants) and non-injectivity; its kinetic properties include the nondegeneracy of its equilibria, and its lack of ACR species\\}
        \label{tab:INRESproperties}
        \begin{tabular}{@{}|c|ll|@{}}
            \hline
            \textbf{Property Class} & \multicolumn{2}{c|}{\textbf{INRES}} \\
            \hline
            \multirow{5}{*}{Network} & Closed & Not (weakly) reversible \\
             & $t$-minimal & Branching \\
             & \multicolumn{2}{l|}{Terminal classes include points and cycles} \\
             & Positive dependent & Deficiency 19 \\
             & Conservative & Discordant \\
            \hline
            Structo-Kinetic & Monostationary & Non-injective \\
            \hline
            Kinetic & Nondegenerate equilibria & No ACR species \\
            \hline
        \end{tabular}
    \end{center}
\end{table}

The complexity of the network is evident in the network's high number of species, complexes, reactant complexes, and reactions, together with its high deficiency ($\delta = 19$). INRES is also a closed network (since $s = 20 < 32 = m$) with high reactant diversity (since $n_r = 41 > 20 = s$).

Viewed as a digraph, the network is not weakly reversible (since $s\ell = 65 \neq 31 = \ell$) but branching (since $r = 44 > 41 = n_r$). Other properties inferred by the numbers include its $t$-minimality (since $t = 31 = \ell$), and non-point terminality and non-cycle terminality (since $t \neq n - n_r = 29$ and $n - n_r \neq 0$, respectively).

The software CRNToolbox \citep{CRNToolbox} can be used to analyze mass action systems, as is the case for INRES. Using the application, we find that INRES is positive dependent, meaning there is a set of rate constants wherein the network has a positive equilibrium. The software's Basic Report also indicates that the network is conservative. The Mass Action Injectivity Report shows that the system is not injective, implying that it is also discordant (this is confirmed by CRNToolbox's Concordance Report). Finally, the Higher Deficiency Report concludes that INRES is monostationary, i.e., for a given set of rate constants, the network cannot admit multiple equilibria (moreover, an equilibrium of INRES cannot be degenerate).

\subsection{The Finest Independent Decomposition}

We now present the finest independent decomposition of $\mathscr{N} = \{R_1, \ldots, R_{44}\}$ (where $R_1, \ldots, R_{44}$ are the reactions of INRES) which will be helpful later in our analyses. Applying the algorithm of Hernandez and De la Cruz \cite{HEDC2021} to determine the finest independent decomposition of a network, we find that this decomposition has 12 subnetworks:

\begin{multicols}{2}
\noindent
\begin{align*}
    & \mathscr{N}_1 = \{ R_1, \ldots, R_7 \} \\
    & \mathscr{N}_2 = \{ R_8, \ldots, R_{14} \} \\
    & \mathscr{N}_3 = \{ R_{15}, R_{16} \} \\
    & \mathscr{N}_4 = \{ R_{17}, \ldots, R_{22} \} \\
    & \mathscr{N}_5 = \{ R_{23}, R_{24}, R_{25} \} \\
    & \mathscr{N}_6 = \{ R_{26}, R_{27} \} \\
    & \mathscr{N}_7 = \{ R_{28}, R_{29}, R_{30} \} \\
    & \mathscr{N}_8 = \{ R_{31}, R_{32} \} \\
    & \mathscr{N}_9 = \{ R_{33}, R_{34} \} \\
    & \mathscr{N}_{10} = \{ R_{35}, R_{36}, R_{37} \} \\
    & \mathscr{N}_{11} = \{ R_{38}, \ldots, R_{42} \} \\
    & \mathscr{N}_{12} = \{ R_{43}, R_{44} \}.
\end{align*}
\end{multicols}

Table \ref{tab:indepDecomp} presents the network numbers of the subnetworks of INRES. The analysis and biological implications of the decomposition are discussed in Section \ref{subsec:differences}.

\begin{table}
    \begin{center}
        \caption{\textbf{Network numbers of the subnetworks of INRES} The finest independent decomposition of INRES has 12 subnetworks representing different processes involved in insulin signaling in type 2 diabetes}
        \label{tab:indepDecomp}
        \begin{tabular}{@{}lcccccccccccc@{}}
            \hline
            \multicolumn{1}{c}{\textbf{Characteristics}} & $\mathscr{N}_1$ & $\mathscr{N}_2$ & $\mathscr{N}_3$ & $\mathscr{N}_4$ & $\mathscr{N}_5$ & $\mathscr{N}_6$ & $\mathscr{N}_7$ & $\mathscr{N}_8$ & $\mathscr{N}_9$ & $\mathscr{N}_{10}$ & $\mathscr{N}_{11}$ & $\mathscr{N}_{12}$ \\
            \hline
            Species & 6 & 6 & 3 & 7 & 4 & 3 & 4 & 3 & 3 & 4 & 5 & 3 \\
            Complexes & 7 & 8 & 4 & 10 & 6 & 4 & 6 & 4 & 4 & 6 & 7 & 4 \\
            Reactant complexes & 5 & 6 & 2 & 6 & 3 & 2 & 3 & 2 & 2 & 3 & 5 & 2 \\
            Reactions & 7 & 7 & 2 & 6 & 3 & 2 & 3 & 2 & 2 & 3 & 5 & 2 \\
            Linkage classes & 2 & 3 & 2 & 4 & 3 & 2 & 3 & 2 & 2 & 3 & 3 & 2 \\
            Strong linkage classes & 5 & 7 & 4 & 10 & 6 & 4 & 6 & 4 & 4 & 6 & 5 & 4 \\
            Terminal strong linkage classes & 2 & 3 & 2 & 4 & 3 & 2 & 3 & 2 & 2 & 3 & 3 & 2 \\
            Rank & 4 & 3 & 1 & 3 & 1 & 1 & 1 & 1 & 1 & 1 & 2 & 1 \\
            Reactant rank & 5 & 6 & 2 & 6 & 3 & 2 & 3 & 2 & 2 & 3 & 5 & 2 \\
            Deficiency & 1 & 2 & 1 & 3 & 2 & 1 & 2 & 1 & 1 & 2 & 2 & 1 \\
            Reactant deficiency & 0 & 0 & 0 & 0 & 0 & 0 & 0 & 0 & 0 & 0 & 0 & 0 \\
            \hline
        \end{tabular}
    \end{center}
\end{table}

\section{Comparative Analysis of Network Properties of \\ INSMS and INRES}
\label{sec:structural}

This section compares the network properties of INSMS and INRES. A \textbf{network} (or structural) \textbf{property} is one that can be specified in terms of the network components alone, i.e., there is no need for specifying a kinetics.

We highlight three differing network properties:

\begin{enumerate}
    \item The sets of species (consequently, functional modules) of the networks are different;
    \item There is a much higher level of complexity in INRES; and
    \item The concordance of INSMS is lost, i.e., INRES is discordant.
\end{enumerate}

We conclude the section with a discussion of further differences between the network properties of the two kinetic realizations.

\subsection{Common Network Properties}
\label{subsec:commonNetwork}

Comparing Tables \ref{tab:INSMSproperties} and \ref{tab:INRESproperties}, we observe that INSMS and INRES are both branching, $t$-minimal, positive dependent, not (weakly) reversible, non-point terminal, and non-cycle terminal. Based on Sections \ref{sec:insms} and \ref{subsec:inres}, both are also closed networks with high reactant diversity.

Lemma 3.5.4 of Feinberg \cite{FEIN2019} and Proposition 1 of Lubenia et al. \cite{LML2022} show that a positive dependent network has a set of rate constants such that the corresponding mass action system has a positive equilibrium. This result is important when we check INRES for ACR species in Section \ref{sec:acr}.

\subsection{Differences in Species Sets and Functional Modules}
\label{subsec:differences}

Among the 20 species in INSMS, only nine appear in INRES as well. Seven of these are involved in the initial signaling steps while the other two are significant in the process of glucose transfer to the glycolytic system. There are 23 unique species in INRES (i.e., those not in INSMS). The presence of these unique species suggests that other complexes get involved when irregularities occur in the insulin signaling pathway, characterizing insulin resistance. Moreover, this difference in species sets clearly shows that the functional modules active in the intermediate steps of the two networks are entirely different. 

Lubenia et al.'s reaction network analysis of INSMS revealed the three functional modules used by Sedaghat et al. in the construction of their model. A coarsening of the finest independent decomposition of INRES also shows these functional modules. A \textbf{coarsening} of a decomposition has some of the original subnetworks combined to form fewer subnetworks. We utilize this concept here to identify the various processes involved in insulin signaling in type 2 diabetes.

Consider the coarsening described in Table \ref{tab:INREScoarseningSedaghat}. $\mathscr{N}_1^*$, $\mathscr{N}_2^*$, and $\mathscr{N}_3^*$ correspond to the subsystems used to construct the Sedaghat et al. model (insulin receptor binding and recycling, postreceptor signaling, and GLUT4 translocation, respectively), albeit with different actors. The last subnetwork $\mathscr{N}_4^*$ appears in the Nyman et al. model only; it is a signaling branch (composed of S6 and S6K formation from mTORC, ERK signaling pathway, and nuclear transcription) that the authors deemed significant in insulin signaling in type 2 diabetes.

\begin{table}[ht]
    \begin{center}
        \caption{\textbf{Coarsening of the finest independent decomposition of INRES in relation to the Sedaghat et al. model} $\mathscr{N}_1^*$, $\mathscr{N}_2^*$, and $\mathscr{N}_3^*$ correspond to the insulin receptor binding and recycling subsystem, postreceptor signaling subsystem, and GLUT4 translocation subsystem, respectively; $\mathscr{N}_4^*$ is a subsystem found in the Nyman et al. model representing a signaling branch (composed of S6 and S6K formation from mTORC, ERK signaling pathway, and nuclear transcription) that the authors deemed significant in insulin signaling in type 2 diabetes\\}
        \label{tab:INREScoarseningSedaghat}
        \begin{tabular}{@{}l|l@{}}
            \multicolumn{2}{c}{$\mathscr{N} = \mathscr{N}_1^* \cup \mathscr{N}_2^* \cup \mathscr{N}_3^* \cup \mathscr{N}_4^*$} \\
            \hline
            $\mathscr{N}_1^*$ & $\mathscr{N}_1$ \\
            $\mathscr{N}_2^*$ & $\mathscr{N}_2 \cup \ldots \cup \mathscr{N}_6$ \\
            $\mathscr{N}_3^*$ & $\mathscr{N}_7 \cup \mathscr{N}_8$ \\
            $\mathscr{N}_4^*$ & $\mathscr{N}_9 \cup \ldots \cup \mathscr{N}_{12}$
        \end{tabular}
    \end{center}
\end{table}

Table \ref{tab:INREScoarseningNyman} presents another coarsening of the 12 subnetworks of INRES which reveals the 9 subsystems that Nyman and colleagues considered in the construction of their model. Subsystems $\mathscr{N}_1^{'}$ to $\mathscr{N}_9^{'}$ correspond to the insulin receptor signaling, IRS1 dynamics, negative feedback to insulin receptors, PKB dynamics, mTORC dynamics, GLUT4 translocation, S6 and S6K formation from mTORC, ERK signaling pathway, and nuclear transcription, respectively.

\begin{table}[ht]
    \begin{center}
        \caption{\textbf{Coarsening of the finest independent decomposition of INRES in relation to the Nyman et al. model} $\mathscr{N}_1^{'}$ to $\mathscr{N}_9^{'}$ correspond to the insulin receptor signaling, IRS1 dynamics, negative feedback to insulin receptors, PKB dynamics, mTORC dynamics, GLUT4 translocation, S6 and S6K formation from mTORC, ERK signaling pathway, and nuclear transcription, respectively\\}
        \label{tab:INREScoarseningNyman}
        \begin{tabular}{@{}l|l@{}}
            \multicolumn{2}{c}{$\mathscr{N} = \mathscr{N}_1^{'} \cup \mathscr{N}_2^{'} \cup \mathscr{N}_3^{'} \cup \mathscr{N}_4^{'} \cup \mathscr{N}_5^{'} \cup \mathscr{N}_6^{'} \cup \mathscr{N}_7^{'} \cup \mathscr{N}_8^{'} \cup \mathscr{N}_9^{'}$} \\
            \hline
            $\mathscr{N}_1^{'}$ & $\mathscr{N}_1$ \\
            $\mathscr{N}_2^{'}$ & $\mathscr{N}_2$ \\
            $\mathscr{N}_3^{'}$ & $\mathscr{N}_3$ \\
            $\mathscr{N}_4^{'}$ & $\mathscr{N}_4$ \\
            $\mathscr{N}_5^{'}$ & $\mathscr{N}_5 \cup \mathscr{N}_6$ \\
            $\mathscr{N}_6^{'}$ & $\mathscr{N}_7 \cup \mathscr{N}_8$ \\
            $\mathscr{N}_7^{'}$ & $\mathscr{N}_9 \cup \mathscr{N}_{10}$ \\
            $\mathscr{N}_8^{'}$ & $\mathscr{N}_{11}$ \\
            $\mathscr{N}_9^{'}$ & $\mathscr{N}_{12}$
        \end{tabular}
    \end{center}
\end{table}

In the finest independent decomposition of INRES, subnetwork $\mathscr{N}_2$ is significant because it contains reaction $R_{11}$ which exhibits the mTORC feedback to IRS1. Affecting this feedback are reactions $R_{23}$ and $R_{24}$ in subnetwork $\mathscr{N}_5$: these are the reactions for the activation of mTORC. These reactions together form the ``single mechanism'' that explains insulin resistance in type 2 diabetes according to Nyman et al.: an attenuated positive feedback from mTORC to IRS1.

These analyses show how decomposition theory aids in identifying the functional modules, and crucial processes and reactions involved in insulin signaling in type 2 diabetes.

Aside from the number of species and functional modules, the number of reactions and the value of the deficiency of INRES also indicate that it is more complex than INSMS. Although INRES has only nine more reactions (44 vs 35), its deficiency is nearly triple that of INSMS (19 vs 7), signifying a much higher level of (linear) dependence among the reactions. On the other hand, the reactant deficiency more than doubles ``only'' from 4 to 9. This appears to be mainly due to the (maximally) high reactant rank of INRES (32 vs 20 for INSMS). The length of the finest independent decomposition of INRES (12 vs 10 for INSMS) gives further testament to its higher complexity compared to INSMS. Although this difference is relatively moderate, this is consistent with the rank relation (20 for INRES vs 15 for INSMS).

\subsection{Concordance vs Discordance}

The most significant difference between INSMS and INRES in terms of network properties is the concordance of the former and the discordance of the latter. Discordance marks the loss of a network characteristic which Shinar and Feinberg \cite{SHFE2013} consider to indicate ``architectures that, by their very nature, enforce duller, more restrictive behavior despite what might be great intricacy in the interplay of many species, even independent of values that kinetic parameters might take''. The discordance of INRES suggests that the interactions between the species in the system is less restrictive, leading to irregularities in the insulin signaling process. The root causes of discordance will be discussed in Section \ref{subsec:injectivity}.

\begin{remark}
    While large classes of biochemical systems have been shown to be discordant, only a handful of systems have been published as having the concordance property. For example, Fari\~{n}as et al. \cite{FML2021} have shown that the independent kinetic realization of any S-system in two or more variables is discordant. To our knowledge, only three concordant biochemical networks have been published to date: a kinetic realization of the Wnt signaling pathway in Chapter 10.5.4 of Feinberg \cite{FEIN2019}, of the insulin signaling in healthy cells by Lubenia et al. \cite{LML2022}, and of the Schmitz carbon cycle model by Fortun and Mendoza \cite{FOME2023}.
\end{remark}

\subsection{Further Differences in Network Properties}

Table \ref{tab:furtherDiff} provides an overview of further differences between INSMS and INRES.

\begin{table}[ht]
    \begin{center}
        \caption{\textbf{Further differences between INSMS and INRES} The nonconservativity of INSMS, and its deficiency zero subnetworks and rank 1 subnetworks are significant in the existence of ACR species in the network; INRES, on the other hand, is conservative and, as shown in Section \ref{sec:acr}, has no ACR species\\}
        \label{tab:furtherDiff}
        \begin{tabular}{@{}|l|l|l|@{}}
            \hline
            \multicolumn{1}{|c|}{\textbf{INSMS}} & \multicolumn{1}{c|}{\textbf{INRES}} & \multicolumn{1}{c|}{\textbf{Comment}} \\
            \hline
            Nonconservative & Conservative & \\
            \hline
            5 (of 10) subnetworks & 0 (of 12) subnetworks & Large subnetwork ($s = 10$) with \\
            with $\delta = 0$ & with $\delta = 0$ & $\delta = 0$ in INSMS \\
            \hline
            9 (of 10) subnetworks  & 8 (of 12) subnetworks & \\
            with $s = 1$ & with $s = 1$ & \\
            \hline
            0 (of 10) subnetworks & 12 (of 12) subnetworks & \\
            with $\delta_{\rho} = 0$ & with $\delta_{\rho} = 0$ & \\
            \hline
        \end{tabular}
    \end{center}
\end{table}

The analysis of Lubenia et al. showed that INSMS is nonconservative while CRNToolbox results reveal that INRES is conservative. The nonconservativity of INSMS was a necessary condition for the ACR in 8 species in INSMS to be inferred from ACR in its deficiency zero subnetwork. Meanwhile, INRES has no deficiency zero subnetwork and, as will be shown in Section \ref{sec:acr}, has no ACR species as well.

Lubenia et al. also observered that INSMS has nine (out of 10) subnetworks whose rank is 1. In INRES, eight (out of 12) subnetworks have rank 1. The Meshkat et al. criterion \citep{MEST2022} was applied to the INSMS rank 1 subnetworks to determine two ACR species. We tried applying the said criterion to rank 1 subnetworks of INRES to determine ACR species but the effort yielded no results.

Finally, another difference between INSMS and INRES is terms of the number of subnetworks with zero reactant deficiency: INSMS has none while every subnetwork of INRES has a reactant deficiency of zero.

\section{Comparative Analysis of Structo-kinetic and \\ Kinetic Properties of INSMS and INRES}
\label{sec:kinetic}

In this section, we discuss the structo-kinetic and kinetic properties of INSMS and INRES, and explore their relationships to insulin signaling in healthy and insulin-resistant cells. \textbf{Structo-kinetic properties} are properties that depend on both network and kinetic properties while \textbf{(purely) kinetic properties} depend on the kinetics alone.

We also highlight in this section the most striking difference in kinetic properties of the two networks: INSMS has 8 ACR species while INRES has none.

\subsection{Common Structo-kinetic Properties}
\label{subsec:commonSK}

The first common structo-kinetic property of INSMS and INRES is the absence of complex balanced equilibria since both networks are not weakly reversible.

Secondly, the $t$-minimality of both networks implies the coincidence of their respective kinetic and stoichiometric subspaces \citep{FEHO1977}.

A notable third structo-kinetic commonality of INSMS and INRES is their monostationarity for any mass action kinetics: the network cannot admit multiple positive equilibria for a given a set of rate constants. For INSMS, this property derives from a general property of weakly monotonic kinetics on a concordant network. For INRES, monostationarity was concluded by the Higher Deficiency Report of CRNToolbox and is considered an infrequent phenomenon in view of the underlying network's discordance.

\subsection{Injective vs Non-injective}
\label{subsec:injectivity}

The reaction network analysis of INSMS showed that it is injective since the network is concordant. On the other hand, the non-injectivity of INRES is shown by the Mass Action Injectivity Report of the CRNToolbox. 

Based on Theorem 4.11 of Shinar and Feinberg \cite{SHFE2012}, the class of concordant networks is precisely the class of networks that are injective for every assignment of a weakly monotonic kinetics. Since every mass action kinetics is weakly monotonic, the root cause of the discordance of INRES is the combination of two factors: the weak monotonicity and non-injectivity of its kinetics.

\subsection{Monostationarity in All Weakly Monotonic Systems vs \\ Multistationarity for Some Weakly Monotonic Systems}

Sections \ref{subsec:commonSK} and \ref{subsec:injectivity} show that while the injectivity of INSMS, induced by a weakly monotonic kinetics on a concordant network, is lost on the discordant INRES, monostationarity is maintained.

Since INSMS is a concordant network, then it has injectivity in all weakly monotonic kinetic systems derived from it (Theorem 4.11 of Shinar and Feinberg \cite{SHFE2012}). Moreover, together with Remark 4.4 in Shinar and Feinberg \cite{SHFE2012}, we conclude that INSMS is monostationary in all weakly monotonic monotonic systems.

Despite its monostationarity for any mass action kinetics, it follows from Theorem 10.5.10 of Feinberg \cite{FEIN2019} that for some weakly monotonic kinetics, INRES will have the capacity for multiple positive equilibria in a stoichiometric class, i.e., the network can display multistationarity. The root of this is, hence, the network property of discordance in combination with the weak monotonicity of its kinetics.

\subsection{Common Kinetic Properties}

We observed only a single common kinetic property of INSMS and INRES: the non-degeneracy of all their positive equilibria. This property can be derived from the $t$-minimality of both networks: $t$-minimality is a necessary condition for the existence of nondegenerate equilibria, as detailed in Remark 4.11 of Feinberg and Horn \cite{FEHO1977}.

\begin{remark}
    \noindent
    \begin{enumerate}
        \item This kinetic property also implies the common structo-kinetic property of coincidence of the kinetic and stoichiometric subsets.
        \item The nondegeneracy of all equilibria of INSMS was a consequence of concordance of the underlying network: if a concordant network has weakly monotonic kinetics, then all equilibria are degenerate. On the other hand, for INRES, the nondegeneracy of its equilibria came from CRNToolbox results.
    \end{enumerate}
\end{remark}

\subsection{Presence of ACR in 8 of 20 Species vs. Absence of ACR in \\ 32 Species}
\label{sec:acr}

\textbf{ACR} is the invariance of the concentrations of a species at all positive equilibria of a kinetic system \citep{SHFE2010, SHFE2011}. In Section \ref{subsec:commonNetwork}, we showed that INSMS and INRES have a positive equilibrium for a set of rate constants. Thus, we can check for ACR species in both networks. Lubenia and colleagues already showed that INSMS has 8 species (out of 20) with ACR and it is noteworthy that this set of species includes GLUT4, the glucose transporter essential for feeding the glycolytic process with glucose. ACR of a species suggests that the species is reliable in maintaining the efficiency of the process it belongs to.

Applying the algorithm of Fontanil et al. \cite{FOMF2021} and the Meshkat et al. criterion \citep{MEST2022}, we find that, remarkably, none of the 32 species in INRES have ACR. Significantly, GLUT4 lost its concentration robustness in the insulin-resistant case, suggesting that it lost its reliability in shuttling glucose to the cell, a crucial process in sustaining efficient energy production.

\begin{remark}
    Aside from GLUT4, five of the other species from INSMS, which lost their ACR in INRES, are pooled insulin receptors. Hence, the first two key biological findings of Nyman et al. on lower concentration of receptors and GLUT4 (see Section \ref{subsec:NymanBiology}) are also reflected in the loss of ACR.
\end{remark}

\subsubsection{Equilibria Parametrization Method Applied to the Sedaghat et al. Model}
\label{sec:param}

Hernandez et al. \cite{HLJK2022} developed a method to determine the parameterized equilibrium of a mass action system. They did this using the fact that the set of positive equilibria of the whole network is the intersection of the equilibria sets of the subnetworks \citep{FEIN1987}, together with network translation methods developed by Johnston and colleagues \citep{JB2019, JMP2019}.

The analysis of INSMS showed that the number of ACR species $m_\mathrm{ACR} \geq 8$ for all rate constants and $m_\mathrm{ACR} = 8$ for some. Subsequently, Hernandez et al. \cite{HLJK2022} sharpened this to $m_\mathrm{ACR} = 8$ for all rate constants using their method applied to the Sedaghat et al. model.

\subsubsection{Equilibria Parametrization Method Applied to the Nyman et al. Model}
\label{sec:eqParam}

To verify that INRES indeed has no ACR species, we tried using the Hernandez et al. method to derive the parametrized equilibrium of the Nyman et al. model. However, since their method did not deliver full results, we used network decomposition to compute manually the equilibria relationships and obtained the following solution where $x_i$ denotes the concentration of species $X_i$:

\begin{align*}
    & x_2 = \frac{k_3}{k_1} x_3 \\
    & x_4 = \frac{k_3 (k_2 + k_1)}{k_1 (k_4 + k_6)} x_3 \\
    & x_6 = \frac{k_4 k_3 (k_2 + k_1)}{k_7 k_1 (k_4 + k_6)} x_3 \\
    & x_9 = \left( \frac{k_{10}}{k_8} + \frac{k_{13} k_{11}}{k_8 (k_{12} + k_{13})} x_{31} \right) \frac{x_{10}}{x_7} \\
    & x_{20} = \frac{k_{32}}{\left( \frac{k_{31} k_{19} k_{17}}{k_{30} (k_{18} + k_{19} x_{33})} x_{10} x_{26} x_{33} x_{34} \right) \left[ \left( 1 + \frac{k_{20} k_{11}}{k_{22} (k_{12} + k_{13})} x_{10} x_{31} \right) \frac{k_{28}}{k_{21}} + \frac{k_{29}}{k_{22}} \right]} x_{21} \\
    & x_{22} = \frac{k_{11}}{k_{12} + k_{13}} x_{10} x_{31} \\
    & x_{23} = \frac{k_9 k_{10}}{k_{14} k_8} \frac{x_{10}}{x_7} + \left( \frac{k_9}{k_8} \frac{1}{x_7} + 1 \right) \frac{k_{13} k_{11}}{k_{14} (k_{12} + k_{13})} x_{10} x_{31} \\
    & x_{24} = \frac{k_{16} k_4 k_3 (k_2 + k_1)}{k_{15} k_5 k_1 (k_4 + k_6)} \frac{x_3}{x_7 x_{10}} \\
    & x_{25} = \frac{k_4 k_3 (k_2 + k_1)}{k_5 k_1 (k_4 + k_6)} \frac{x_3}{x_7} \\
    & x_{27} = \frac{k_{17}}{k_{18} + k_{19} x_{33}} x_{10} x_{26} \\
    & x_{28} = \frac{k_{19} k_{17}}{k_{22} (k_{18} + k_{19} x_{33})} x_{10} x_{26} x_{33} \\
    & x_{29} = \left( 1 + \frac{k_{20} k_{11}}{k_{22} (k_{12} + k_{13})} x_{10} x_{31} \right) \frac{k_{19} k_{17}}{k_{21} (k_{18} + k_{19} x_{33})} x_{10} x_{26} x_{33} \\
    & x_{30} = \frac{k_{25}}{\left( \frac{k_{17}}{k_{18} + k_{19} x_{33}} x_{10} x_{26} \right) \left[ \left( 1 + \frac{k_{20} k_{11}}{k_{22} (k_{12} + k_{13})} x_{10} x_{31} \right) \frac{k_{23} k_{19}}{k_{21}} x_{33} + k_{24} \right]} x_{31} \\
    & x_{32} = \frac{k_{27}}{k_{26}} \frac{x_{33}}{x_7} \\
    & x_{35} = \left( \frac{k_{19} k_{17}}{k_{30} (k_{18} + k_{19} x_{33})} x_{10} x_{26} x_{33} x_{34} \right) \left[ \left( 1 + \frac{k_{20} k_{11}}{k_{22} (k_{12} + k_{13})} x_{10} x_{31} \right) \frac{k_{28}}{k_{21}} + \frac{k_{29}}{k_{22}} \right] \\
    & x_{37} = \frac{k_{33}}{k_{34}} x_{31} x_{36} \\
    & x_{39} = \frac{k_{35} k_{33}}{k_{37} k_{34}} x_{31} x_{36} x_{38} + \frac{k_{36}}{k_{37}} x_{38}x_{41} \\
    & x_{40} = \frac{k_{41}}{k_{38} x_7 + \frac{k_{39} k_{11}}{k_{12} + k_{13}} x_{10} x_{31} + k_{40}} x_{41} \\
    & x_{42} = \frac{k_{41}}{k_{42}} x_{41} \\
    & x_{44} = \frac{k_{43}}{k_{44}} x_{41} x_{43}
\end{align*}

Observe that the parametrized equilibrium of the species above depend on other species $x_3$, $x_7$, $x_{10}$, $x_{21}$, $x_{26}$, $x_{31}$, $x_{33}$, $x_{34}$, $x_{36}$, $x_{38}$, $x_{41}$, and $x_{43}$, implying that there are infinitely many possible equilibrium solutions for the system. We, therefore, conclude that INRES has no ACR species.

The parametrized equilibria can potentially be used in collaboration with biologists. For example, let us focus on the equilibrium parametrization of the concentration of the intracellular glucose transporter GLUT4 ($x_{20}$). Measurements of the amount of GLUT4 in experimental data were used by Sedaghat et al. \cite{SSQ2002} and Nyman et al. \cite{NRFBCS2014} in the construction of their models of insulin signaling in healthy and insulin-resistant cells. The ACR of GLUT4 in the healthy cell model suggests that keeping the amount of GLUT4 (approximately) constant could be valuable against insulin resistance as glucose can be efficiently transported to the cell. The equilibria parametrization shows that the concentration of GLUT4 is dependent on the concentrations of other species ($x_{10}$, $x_{21}$, $x_{26}$, $x_{31}$, $x_{33}$, and $x_{34}$). One can possibly work with biologists to determine if the concentration of these species can be modified (e.g., using medical interventions) to ensure that the amount of GLUT4 remains approximately constant or at least bounded, resulting in continued efficient glucose translocation into the cell.

Another possible application of our analysis in relation to ACR is the usage of a network's finest independent decomposition. The idea is to start by looking at the subnetwork/s where a species of interest is present and finding a way to establish ACR of the species in the small subnetwork. One can then try to extend it to bigger subnetworks and, eventually, to the entire network. Working with biologists is, of course, essential in this synthetic biology approach to ensure that results are biologically sound.

\section{Summary and Conclusion}
\label{sec:summary}

To study and understand the underlying mechanism of insulin signaling, we have constructed, analyzed, and compared INSMS and INRES. Section \ref{sec:insms} summarizes the reaction network analysis of Lubenia et al. on a kinetic realization of a model of metabolic insulin signaling in healthy cells (INSMS). In Section \ref{sec:inres}, we constructed a CRN of a model of insulin signaling in an insulin-resistant cell (INRES). 

We showed how to use decomposition theory to identify the subsystems involved in insulin signaling. Subnetworks of INRES revealed the functional modules that are also present in INSMS: insulin receptor binding and recycling, postreceptor signaling, and GLUT4 translocation. Furthermore, additional subsystems are present in INRES but not in INSMS: the S6 and S6K formation from mTORC, ERK signaling pathway, and nuclear transcription. More notably, decomposition theory allowed us to identify the subnetworks of INRES which correspond to the ``single mechanism'' that explains insulin resistance in type 2 diabetes according to Nyman et al.: an attenuated positive feedback from mTORC to IRS1.

We then compared the different network, structo-kinetic, and kinetic properties of the two reaction networks. We have summarized in a table in Appendix C all the properties of INSMS and INRES discussed in this paper. INSMS and INRES are similar with respect to some properties. But more importantly, we highlighted that kinetic system differences revealed notable biological differences of the same system under differing states, i.e., healthy and insulin-resistant.

First, the set of species involved in insulin signaling (and consequently the functional modules they constitute) are very different. This variation in the number of species involved in INSMS and INRES suggests that when irregularities occur in the insulin signaling pathway, other complexes (and, hence, other processes) get involved, characterizing insulin resistance.

Second, the concordance of INSMS is lost in INRES. This suggests less restrictive interplay between the species involved in insulin signaling, leading to unusual activities in the signaling cascade.

Finally, INSMS has 8 ACR species while INRES has none. This marked the loss of ACR in 6 of its common species with INSMS, including the critical GLUT4 which is responsible for transporting glucose into the glycolytic system. The loss of ACR in GLUT4 suggests that, in an insulin-resistant cell, the transporter has lost its reliability in shuttling glucose to the cell, inhibiting efficient cellular energy production. We also talked about possible ways of using information gained from a network's equilibria parametrization and its finest independent decomposition to potentially establish ACR in a species. We emphasized that consultation with biologist is essential to ensure proper biological interpretation of results.

These findings, from our view, have shown the usefulness of doing a kinetic realization analysis, beyond what we already know from the analysis of the dynamical systems of Sedaghat et al. and Nyman et al.



\begin{appendices}

\counterwithin{example}{section}

\section{Notations and Definition of Terms}
\label{app:crnt}

In this section, we lay the foundation of Chemical Reaction Network Theory by discussing the definition of terms used in the paper. After discussing the fundamentals of chemical reaction networks and kinetic systems, we review important terminologies related to decomposition theory.

\subsection{Chemical Reaction Networks}

A \textbf{chemical reaction network} (CRN) $\mathscr{N}$ is a triple $(\mathscr{S}, \mathscr{C}, \mathscr{R})$ of nonempty finite sets $\mathscr{S}$, $\mathscr{C}$, and $\mathscr{R}$ of $m$ species, $n$ complexes, and $r$ reactions, respectively. 


In a CRN, we denote the species as $X_1, \ldots, X_m$. This way, $X_i$ can be identified with the vector in $\mathbb{R}^m$ with 1 in the $i$th coordinate and zero elsewhere. We denote the reactions as $R_1, \ldots, R_r$. We denote the complexes as $C_1, \ldots, C_n$ where the manner in which the complexes are numbered play no essential role. A complex $C_i \in \mathscr{C}$ is given as $\displaystyle C_i = \sum_{j=1}^m c_{ij} X_j$ or as the vector $(c_{i1}, \ldots, c_{im}) \in \mathbb{R}_{\geq 0}^m$ (the subscript $\geq 0$ means we consider only the nonnegative real numbers). The coefficient $c_{ij}$ is called the \textbf{stoichiometric coefficient} of species $X_j$ in complex $C_i$. Stoichiometric coefficients are all nonnegative numbers. We define the \textbf{zero complex} as the zero vector in $\mathbb{R}^m$. The ordered pair $(C_i, C_j)$ corresponds to the familiar notation $C_i \rightarrow C_j$ which indicates the reaction where complex $C_i$ reacts to complex $C_j$. We call $C_i$ the \textbf{reactant complex} and $C_j$ the \textbf{product complex}. We denote the number of reactant complexes as $n_r$.


\begin{example}
Consider the reaction
\begin{equation*}
    2 X_1 + X_2 \rightarrow 2 X_3.
\end{equation*}
$X_1$, $X_2$, and $X_3$ are the species, the reactant complex is $2 X_1 + X_2$, and $2 X_3$ is the product complex. The stoichiometric coefficients are 2, 1, and 2 for $X_1$, $X_2$, and $X_3$, respectively.
\end{example}

Let $\mathscr{N} = (\mathscr{S}, \mathscr{C}, \mathscr{R})$ be a CRN. For each reaction $C_i \rightarrow C_j \in \mathscr{R}$, we associate the \textbf{reaction vector} $C_j - C_i \in \mathbb{R}^m$. The linear subspace of $\mathbb{R}^m$ spanned by the reaction vectors is called the \textbf{stoichiometric subspace} of $\mathscr{N}$, defined as $S = \text{span}\{C_j - C_i \in \mathbb{R}^m \mid C_i \rightarrow C_j \in \mathscr{R}\}$. The \textbf{rank} of $\mathscr{N}$ is given by $s = \text{dim}(S)$, i.e., the rank of the network is the rank of its set of reaction vectors. In this paper, we sometimes use the notation $\mathscr{N} = \{R_1, \ldots, R_r\}$ where we loosely use the notation $R_i$ to refer to either reaction $i$ or its corresponding reaction vectors.

Two vectors $x^*, x^{**} \in \mathbb{R}^m$ are said to be \textbf{stoichiometrically compatible} if $x^* - x^{**}$ is an element of the stoichiometric subspace $S$. Stoichiometric compatibility is an equivalence relation that induces a partition of $\mathbb{R}_{\geq 0}^m$ or $\mathbb{R}_{>0}^m$ into equivalence classes called the \textbf{stoichiometric compatibility classes} or \textbf{positive stoichiometric compatibility classes}, respectively, of the network. In particular, the stoichiometric compatibility class containing $x \in \mathbb{R}_{\geq 0}^m$ is the set $(x + S) \cap \mathbb{R}_{\geq 0}^m$ where $x + S$ is the left coset of $S$ containing $x$. Similarly, the positive stoichiometric compatibility class containing $x \in \mathbb{R}_{>0}^m$ is the set $(x + S) \cap \mathbb{R}_{>0}^m$.

The \textbf{molecularity matrix} $Y$ is an $m \times n$ matrix whose entry $Y_{ij}$ is the stoichiometric coefficient of species $X_i$ in complex $C_j$. The \textbf{incidence matrix} $I_a$ is an $n \times r$ matrix whose entry $(I_a)_{ij}$ is defined as follows:
\begin{equation*}
    (I_a)_{ij} =
    \left\{
        \begin{array}{rl}
            -1	& \text{if $C_i$ is the reactant complex of reaction $R_j$} \\
            1	& \text{if $C_i$ is the product complex of reaction $R_j$} \\
            0	& \text{otherwise} \\
        \end{array}
    \right..
\end{equation*}
The \textbf{stoichiometric matrix} $N$ is the $m \times r$ matrix given by $N = Y I_a$. The columns of $N$ are the reaction vectors of the system. From the definition of stoichiometric subspace, we can see that $S$ is the image of $N$, written as $S = \text{Im}(N)$. Observe that $s = \text{dim}(S) = \text{dim}(\text{Im}(N)) = \text{rank}(N)$.

\begin{example}
\label{ex:2}
Consider the following CRN:
\begin{align*}
    & R_1: 2 X_1 \rightarrow X_3 \\
    & R_2: X_2 + X_3 \rightarrow X_3 \\
    & R_3: X_3 \rightarrow X_2 + X_3 \\
    & R_4: 3 X_4 \rightarrow X_2 + X_3 \\
    & R_5: 2 X_1 \rightarrow 3 X_4.
\end{align*}
The set of species and complexes are $\mathscr{S} = \{ X_1, X_2, X_3, X_4 \}$ and $\mathscr{C} = \{ 2 X_1, X_2 + X_3, X_3, 3 X_4 \}$, respectively. Thus, there are $m = 4$ species, $n = 4$ complexes, $n_r = 4$ reactant complexes, and $r = 5$ reactions. The network's molecularity matrix, incidence matrix, and stoichiometric matrix are as follows:
\begin{multicols}{2}
    \begin{center}
        \[ Y =
            \begin{blockarray}{ccccc}
                2 X_1 & X_2 + X_3 & X_3 & 3 X_4	\\
                \begin{block}{[cccc]l}
                    2 & 0 & 0 & 0 \topstrut & X_1	\\
                    0 & 1 & 0 & 0 & X_2	\\
                    0 & 1 & 1 & 0 & X_3 \\
                    0 & 0 & 0 & 3 \botstrut	& X_4	\\
                \end{block}
            \end{blockarray}
        \]
    \end{center}
    \begin{center}
        \[ I_a =
            \begin{blockarray}{rrrrrr}
                R_1 & R_2 & R_3 & R_4 & R_5\textcolor{white}{1}	\\
                \begin{block}{[rrrrr]l}
                    \textcolor{white}{1}-1 & 0 & 0 & 0 & -1\textcolor{white}{1} \topstrut & 2 X_1	\\
                    0 & -1 & 1 & 1 & 0\textcolor{white}{1} & X_2 + X_3	\\
                    1 & 1 & -1 & 0 & 0\textcolor{white}{1} & X_3 \\
                    0 & 0 & 0 & -1 & 1\textcolor{white}{1} \botstrut	& 3 X_4	\\
                \end{block}
            \end{blockarray}
        \]
    \end{center}
\end{multicols}
\begin{equation*}
    N = Y I_a =
        \left[
            \begin{array}{rrrrr}
                -2 & 0 & 0 & 0 & -2 \\
                0 & -1 & \textcolor{white}{-}1 & 1 & 0 \\
                1 & 0 & 0 & 1 & 0 \\
                0 & 0 & 0 & -3 & 3 \\
            \end{array}
        \right].
\end{equation*}
The network has rank $s = \text{rank}(N) = 3$.
\end{example}

CRNs can be viewed as directed graphs where the complexes are represented by vertices and the reactions by edges. The \textbf{linkage classes} of a CRN are the subnetworks of its reaction graph where for any complexes $C_i$ and $C_j$ of the subnetwork, there is a path between them. The number of linkage classes is denoted by $\ell$. The linkage class is said to be a \textbf{strong linkage class} if there is a directed path from $C_i$ to $C_j$, and vice versa, for any complexes $C_i$ and $C_j$ of the subnetwork. The number of strong linkage classes is denoted by $s \ell$. Moreover, \textbf{terminal strong linkage classes}, the number of which is denoted as $t$, are the maximal strongly connected subnetworks where there are no edges (reactions) from a complex in the subgraph to a complex outside the subnetwork. Complexes belonging to terminal strong linkage classes are called \textbf{terminal}; otherwise, they are called \textbf{nonterminal}.

In Example \ref{ex:2}, the number of linkage classes is $\ell = 1$: $\{2 X_1, X_3, X_2 + X_3, 3 X_4\}$; the number of strong linkage classes is $s \ell = 3$: $\{X_3, X_2 + X_3\}, \{2 X_1\}, \{3 X_4\}$; and the number of terminal strong linkage classes is $t = 1$: $\{X_3, X_2 + X_3\}$. $X_3$ and $X_2 + X_3$ are terminal complexes while $2 X_1$ and $3 X_4$ are nonterminal complexes.

A CRN is called \textbf{weakly reversible} if $s \ell = \ell$, \textbf{$t$-minimal} if $t = \ell$, \textbf{point terminal} if $t = n - n_r$, and \textbf{cycle terminal} if $n - n_r = 0$. The \textbf{deficiency} of a CRN is given by $\delta = n - \ell - s$.

For a CRN $\mathscr{N}$, the linear subspace of $\mathbb{R}^m$ generated by the reactant complexes is called the \textbf{reactant subspace} of $\mathscr{N}$, defined as $R = \text{span}\{C_i \in \mathbb{R}^m \mid C_i \rightarrow C_j \in \mathscr{R}\}$. The \textbf{reactant rank} of $\mathscr{N}$ is given by $q = \text{dim}(R)$, i.e., the reactant rank of the network is the rank of its set of complexes. The \textbf{reactant deficiency} of $\mathscr{N}$ is given by $\delta_p = n_r - q$.

To make sense of the reactant subspace $R$, write the incidence matrix as $I_a = I_a^+ - I_a^-$ where $I_a^+$ consists only of the 0's and 1's in $I_a$ while $I_a^-$ contains only the 0's and absolute values of the $-1$'s. We form the \textbf{reactant matrix} $N^-$ (size $m \times r$) given by $N^- = Y I_a^-$. The columns of $N^-$ contains the reactant complexes of the system. From the definition of reactant subspace, we can see that $R$ is the image of $N^-$, written as $R = \text{Im}(N^-)$. Observe that $q = \text{dim}(R) = \text{dim}(\text{Im}(N^-)) = \text{rank}(N^-)$.

The incidence matrix of the network in Example \ref{ex:2} can be written as
\begin{align*}
    I_a &= I_a^+ - I_a^- \\
    \left[
        \begin{array}{rrrrr}
            -1 & 0 & 0 & 0 & -1 \\
            0 & -1 & 1 & 1 & 0 \\
            1 & 1 & -1 & 0 & 0 \\
            0 & 0 & 0 & -1 & 1 \\
        \end{array}
    \right]
    &=
    \left[
        \begin{array}{rrrrr}
            0 & 0 & 0 & 0 & 0 \\
            0 & 0 & 1 & 1 & 0 \\
            1 & 1 & 0 & 0 & 0 \\
            0 & 0 & 0 & 0 & 1 \\
        \end{array}
    \right]
    -
    \left[
        \begin{array}{rrrrr}
            1 & 0 & 0 & 0 & 1 \\
            0 & 1 & 0 & 0 & 0 \\
            0 & 0 & 1 & 0 & 0 \\
            0 & 0 & 0 & 1 & 0 \\
        \end{array}
    \right]
\end{align*}
allowing us to form the reactant matrix
\begin{equation*}
    N^- = Y I_a^- =
    \left[
        \begin{array}{rrrrr}
            2 & 0 & 0 & 0 & 2 \\
            0 & 1 & 0 & 0 & 0 \\
            0 & 1 & 1 & 0 & 0 \\
            0 & 0 & 0 & 3 & 0 \\
        \end{array}
    \right].
\end{equation*}
The network has reactant rank $q = \text{rank}(N^-) = 4$ and reactant deficiency $\delta_p = n_r - q = 4 - 4 = 0$.

\subsection{Chemical Kinetic Systems}

A \textbf{kinetics} $K$ for a CRN $\mathscr{N} = (\mathscr{S}, \mathscr{C}, \mathscr{R})$ is an assignment to each reaction $C_i \rightarrow C_j \in \mathscr{R}$ of a rate function $K_{C_i \rightarrow C_j}: \mathbb{R}_{\geq 0}^m \rightarrow \mathbb{R}_{\geq 0}$. 
The system $(\mathscr{N}, K)$ is called a \textbf{chemical kinetic system} (CKS). 

A kinetics gives rise to two closely related objects: the species formation rate function and the associated ordinary differential equation system.

The \textbf{species formation rate function} (SFRF) of a CKS is given by
\begin{equation*}
    f(x) = \sum_{C_i \rightarrow C_j} K_{C_i \rightarrow C_j} (x) (C_j - C_i)
\end{equation*}
where $x$ is the vector of concentrations of species in $\mathscr{S}$ and $K_{C_i \rightarrow C_j}$ is the rate function assigned to reaction $C_i \rightarrow C_j \in \mathscr{R}$. The SFRF is simply the summation of the reaction vectors for the network, each multiplied by the corresponding rate function. The \textbf{kinetic subspace} $\mathcal{K}$ for a CKS is the linear subspace of $\mathbb{R}^m$ defined by $\mathcal{K} = \text{span}\{ \text{Im}(f) \}.$ Note that the SFRF can be written as $f(x) = N K(x)$ where $K$ the vector of rate functions. The equation $\dot{x} = f(x)$ is the \textbf{ordinary differential equation} (ODE) \textbf{system} or \textbf{dynamical system} of the CKS.

The ODE system of the CRN in Example \ref{ex:2} can be written as
\begin{equation*}
    \dot{x} =
    \left[
        \begin{array}{c}
            \dot{x}_1 \\
            \dot{x}_2 \\
            \dot{x}_3 \\
            \dot{x}_4 \\
        \end{array}
    \right]
    =
    \left[
        \begin{array}{rrrrr}
            -2 & 0 & 0 & 0 & -2 \\
            0 & -1 & \textcolor{white}{-}1 & 1 & 0 \\
            1 & 0 & 0 & 1 & 0 \\
            0 & 0 & 0 & -3 & 3 \\
        \end{array}
    \right]
    \left[
        \begin{array}{l}
            k_1 x_1^{f_{11}} \\
            k_2 x_2^{f_{22}} x_3^{f_{23}} \\
            k_3 x_3^{f_{33}} \\
            k_4 x_4^{f_{44}} \\
            k_5 x_1^{f_{51}} \\
        \end{array}
    \right]
    = N K(x).
\end{equation*}

A zero of the SFRF is called an \textbf{equilibrium} or a \textbf{steady state} of the system. If $f$ is differentiable, an equilibrium $x^*$ is called \textbf{degenerate} if $\text{Ker}(J_{x^*} (f)) \cap S \neq \{0\}$ where $J_{x^*} (f)$ is the Jacobian of $f$ evaluated at $x^*$ and Ker is the kernel function; otherwise, the equilibrium is said to be \textbf{nondegenerate}.

A vector $x \in \mathbb{R}_{>0}^m$ is called \textbf{complex balanced} if $K(x)$ is contained in $\text{Ker}(I_a)$ where $I_a$ is the incidence matrix. Furthermore, if $x$ is a positive equilibrium (i.e., all coordinates are positive), then we call it a \textbf{complex balanced equilibrium}. A CKS is called \textbf{complex balanced} if it has a complex balanced equilibrium.

The reaction vectors of a CRN are \textbf{positively dependent} if, for each reaction $C_i \rightarrow C_j \in \mathscr{R}$, there exists a positive number $\alpha_{C_i \rightarrow C_j}$ such that
\begin{equation*}
    \sum_{C_i \rightarrow C_j} \alpha_{C_i \rightarrow C_j} (C_j - C_i) = 0.
\end{equation*}
A CRN with positively dependent reaction vectors is said to be \textbf{positive dependent}. Shinar and Feinberg \cite{SHFE2012} showed that a CKS can admit a positive equilibrium only if its reaction vectors are positively dependent. The \textbf{set of positive equilibria} of a CKS is given by
\begin{equation*}
    E_+ (\mathscr{N}, K) = \{x \in \mathbb{R}_{>0}^m \mid f(x) = 0\}.
\end{equation*}
A CRN is said to \textbf{admit multiple (positive) equilibria} if there exist positive rate constants such that the ODE system admits more than one stoichiometrically compatible equilibria. Analogously, the \textbf{set of complex balanced equilibria} of a CKS $(\mathscr{N}, K)$ is given by
\begin{equation*}
    Z_+ (\mathscr{N}, K) = \{ x \in \mathbb{R}_{>0}^m \mid I_a K(x) = 0 \} \subseteq E_+ (\mathscr{N}, K).
\end{equation*}

Let $F$ be an $r \times m$ matrix of real numbers. Define $x^F$ by $\displaystyle (x^F)_i = \prod_{j=1}^m x_j^{f_{ij}}$ for $i = 1, \ldots, r$. A \textbf{power law kinetics} (PLK) assigns to each $i$th reaction a function
\begin{equation*}
    K_i (x) = k_i (x^F)_i
\end{equation*}
with \textbf{rate constant} $k_i > 0$ and \textbf{kinetic order} $f_{ij} \in \mathbb{R}$. The vector $k \in \mathbb{R}^r$ is called the \textbf{rate vector} and the matrix $F$ is called the \textbf{kinetic order matrix}. We refer to a CRN with PLK as a \textbf{power law system}. The PLK becomes the well-known \textbf{mass action kinetics} (MAK) if the kinetic order matrix consists of stoichiometric coefficients of the reactants. We refer to a CRN with MAK as a \textbf{mass action system}.

In the ODE system of Example \ref{ex:2}, we assumed PLK so that the kinetic order matrix is
\begin{equation*}
    F =
    \left[
        \begin{array}{cccc}
            f_{11} & 0 & 0 & 0 \\
            0 & f_{22} & f_{23} & 0 \\
            0 & 0 & f_{33} & 0 \\
            0 & 0 & 0 & f_{44} \\
            f_{51} & 0 & 0 & 0 \\
        \end{array}
    \right]
\end{equation*}
where $f_{ij} \in \mathbb{R}$. If we assume MAK, the kinetic order matrix is
\begin{equation*}
    F =
    \left[
        \begin{array}{cccc}
            2 & 0 & 0 & 0 \\
            0 & 1 & 1 & 0 \\
            0 & 0 & 1 & 0 \\
            0 & 0 & 0 & 3 \\
            2 & 0 & 0 & 0 \\
        \end{array}
    \right].
\end{equation*}

The reactions $R_i, R_j \in \mathscr{R}$ are called \textbf{branching reactions} if they have the same reactant complex. One way to check if we have identified all branching reactions is through the formula
\begin{equation*}
    r - n_r = \sum_{C_i} (\vert R_{C_i} \vert - 1)
\end{equation*}
where $C_i$ is the reactant complex in $C_i \rightarrow C_j \in \mathscr{R}$, $R_{C_i}$ is the set of branching reactions of $C_i$, and $\vert R_{C_i} \vert$ is the cardinality of $R_{C_i}$. $r - n_r = 0$ if and only if all reactant complexes are nonbranching. A CRN is called \textbf{branching} if $r > n_r$.


Note that the stoichiometric subspace $S$ is just the set of all linear combinations of the reaction vectors, i.e., the set of all vectors in $\mathbb{R}^m$ can be written in the form
\begin{equation*}
    \sum_{C_i \rightarrow C_j} \alpha_{C_i \rightarrow C_j} (C_j - C_i).
\end{equation*}

Let $L: \mathbb{R}^r \rightarrow S$ be the linear map defined by
\begin{equation*}
    L(\alpha) = \sum_{C_i \rightarrow C_j} \alpha_{C_i \rightarrow C_j} (C_j - C_i).
\end{equation*}
$\text{Ker}(L)$ is the set of all vectors $\alpha \in \mathbb{R}^r$ such that $L(\alpha) = 0$.

We define the \textbf{support} of complex $C_i \in \mathscr{C}$ is $\text{supp}(C_i) = \{X_j \in \mathscr{S} \mid c_{ij} \neq 0\}$, i.e, it is the set of all species that have nonzero stoichiometric coefficients in complex $C_i$. Alternatively, we can also define the support of a vector $x \in \mathbb{R}^n$ as the set of indices of the coordinates of $x$ from the index set $\{ 1, \dots, n \}$ for which $x_i \neq 0$.

We say that a CRN is \textbf{concordant} if there do not exist an $\alpha \in \text{Ker}(L)$ and a nonzero $\sigma \in S$ having the following properties:
\begin{enumerate}[($i$)]
    \item For each $C_i \rightarrow C_j \in \mathscr{R}$ such that $\alpha_{C_i \rightarrow C_j} \neq 0$, $\text{supp}(C_i)$ contains a species $X$ for which $\text{sgn}(\sigma_X) = \text{sgn}(\alpha_{C_i \rightarrow C_j})$ where $\sigma_X$ denotes the term in $\sigma$ involving the species $X$ and $\text{sgn}(\cdot)$ is the signum function.
    \item For each $C_i \rightarrow C_j \in \mathscr{R}$ such that $\alpha_{C_i \rightarrow C_j} = 0$, either $\sigma_X = 0$ for all $X \in \text{supp}(C_i)$, or else $\text{supp}(C_i)$ contains species $X$ and $X'$ for which $\text{sgn}(\sigma_X) = -\text{sgn}(\sigma_{X'})$, but not zero.
\end{enumerate}
A network that is not concordant is \textbf{discordant}.

A CKS is \textbf{injective} if, for each pair of distinct stoichiometrically compatible vectors $x^*, x^{**} \in \mathbb{R}_{\geq 0}^m$, at least one of which is positive,
\begin{equation*}
    \sum_{C_i \rightarrow C_j} K_{C_i \rightarrow C_j} (x^{**}) (C_j - C_i) \neq \sum_{C_i \rightarrow C_j} K_{C_i \rightarrow C_j} (x^*) (C_j - C_i).
\end{equation*}
Clearly, an injective kinetic system cannot admit two distinct stoichiometrically compatible equilibria, at least one of which is positive.

A kinetics for a CRN is \textbf{weakly monotonic} if, for each pair of vectors $x^*, x^{**} \in \mathbb{R}_{\geq 0}^m$, the following implications hold for each reaction $C_i \rightarrow C_j \in \mathscr{R}$ such that $\text{supp}(C_i) \subset \text{supp}(x^*)$ and $\text{supp}(C_i) \subset \text{supp}(x^{**})$:
\begin{enumerate}[($i$)]
    \item $K_{C_i \rightarrow C_j} (x^{**}) > K_{C_i \rightarrow C_j} (x^*)$ implies that there is a species $X_k \in \text{supp}(C_i)$ with $x_k^{**} > x_k^*$.
    \item $K_{C_i \rightarrow C_j} (x^{**}) = K_{C_i \rightarrow C_j} (x^*)$ implies that $x_k^{**} = x_k^*$ for all $X_k \in \text{supp}(C_i)$ or else there are species $X_k, X_k' \in \text{supp}(C_i)$ with $x_k^{**} > x_k^*$ and $(x_k')^{**} < (x_k')^*$.
\end{enumerate}
We say that a CKS is \textbf{weakly monotonic} when its kinetics is weakly monotonic.

\begin{example}
    Every MAK is weakly monotonic.
\end{example}

\subsection{Decomposition Theory}

A \textbf{covering} of a CRN is a collection of subsets $\{\mathscr{R}_1, \ldots, \mathscr{R}_k\}$ whose union is $\mathscr{R}$. A covering is called a \textbf{decomposition} of $\mathscr{N}$ if the sets $\mathscr{R}_i$ form a partition of $\mathscr{R}$. $\mathscr{R}_i$ defines a subnetwork $\mathscr{N}_i$ of $\mathscr{N}$ where $\mathscr{N}_i = (\mathscr{S}_i, \mathscr{C}_i, \mathscr{R}_i)$ such that $\mathscr{C}_i $ consists of all complexes occurring in $\mathscr{R}_i$ and $\mathscr{S}_i$ has all the species occurring in $\mathscr{C}_i$. In this paper, we will denote a decomposition as a union of the subnetworks: $\mathscr{N} = \mathscr{N}_1 \cup \ldots \cup \mathscr{N}_k$. We refer to a ``decomposition'' with a single ``subnetwork'' $\mathscr{N} = \mathscr{N}_1$ as the \textbf{trivial decomposition}. Furthermore, when a network has been decomposed into subnetworks, we can refer to the said network as the \textbf{parent network}.


A decomposition $\mathscr{N} = \mathscr{N}_1 \cup \ldots \cup \mathscr{N}_k$ is \textbf{independent} if the parent network's stoichiometric subspace $S$ is the direct sum of the subnetworks' stoichiometric subspaces $S_i$. Equivalently, the sum is direct when the rank of the parent network is equal to the sum of the ranks of the individual subnetworks, i.e.,
\begin{equation*}
    s = \sum_{i=1}^k s_i \text{ where } s_i = \text{dim}(S_i).
\end{equation*}

A network decomposition $\mathscr{N} = \mathscr{N}_1 \cup \ldots \cup \mathscr{N}_k$ is a \textbf{refinement} of $\mathscr{N} = \mathscr{N}_1' \cup \ldots \cup \mathscr{N}_{k'}'$ (and the latter a \textbf{coarsening} of the former) if it is induced by a refinement $\{\mathscr{R}_1, \ldots, \mathscr{R}_k\}$ of $\{\mathscr{R}_1', \ldots, \mathscr{R}_{k'}'\}$.

\begin{example}
If $\mathscr{N} = \mathscr{N}_1 \cup \mathscr{N}_2 \cup \mathscr{N}_3$ and $\mathscr{N}' = \mathscr{N}_2 \cup \mathscr{N}_3$, then
\begin{itemize}
    \item $\mathscr{N} = \mathscr{N}_1 \cup \mathscr{N}_2 \cup \mathscr{N}_3$ is a refinement of $\mathscr{N} = \mathscr{N}_1 \cup \mathscr{N}'$; and
    \item $\mathscr{N} = \mathscr{N}_1 \cup \mathscr{N}'$ is a coarsening of $\mathscr{N} = \mathscr{N}_1 \cup \mathscr{N}_2 \cup \mathscr{N}_3$.
    \end{itemize}
\end{example}

\section{The Nyman et al. Model}
\label{app:nyman}

This section presents the system of ODEs of the insulin signaling in type 2 diabetes model of Nyman et al. \cite{NRFBCS2014}. It also lists in detail the definition of all the variables involved.

The system of ODEs of the Nyman et al. model are as follows:
\begin{align*}
    & \dot{x}_2 = -k_1 x_2 - k_2 x_2 + k_6 x_4 + k_7 x_6 \\
    & \dot{x}_4 = k_2 x_2 + k_3 x_3 - k_4 x_4 - k_6 x_4 \\
    & \dot{x}_3 = k_1 x_2 - k_3 x_3 \\
    & \dot{x}_7 = k_4 x_4 - k_5 x_7 x_{25} \\
    & \dot{x}_6 = k_5 x_7 x_{25} - k_7 x_6 \\
    & \dot{x}_9 = -k_8 x_7 x_9 - k_9 x_9 + k_{10} x_{10} + k_{14} x_{23} \\
    & \dot{x}_{10} = k_8 x_7 x_9 - k_{10} x_{10} - k_{11} x_{10} x_{31} + k_{12} x_{22} \\
    & \dot{x}_{22} = k_{11} x_{10} x_{31} - k_{12} x_{22} - k_{13} x_{22} \\
    & \dot{x}_{23} = k_9 x_9 + k_{13} x_{22} - k_{14} x_{23} \\
    & \dot{x}_{24} = -k_{15} x_{10} x_{24} + k_{16} x_{25} \\
    & \dot{x}_{25} = k_{15} x_{10} x_{24} - k_{16} x_{25} \\
    & \dot{x}_{26} = -k_{17} x_{10} x_{26} + k_{18} x_{27} + k_{22} x_{28} \\
    & \dot{x}_{27} = k_{17} x_{10} x_{26} - k_{18} x_{27} - k_{19} x_{27} x_{33} \\
    & \dot{x}_{28} = -k_{20} x_{22} x_{28} + k_{21} x_{29} - k_{22} x_{28} \\
    & \dot{x}_{29} = k_{19} x_{27} x_{33} + k_{20} x_{22} x_{28} - k_{21} x_{29} \\
    & \dot{x}_{30} = -k_{23} x_{29} x_{30} - k_{24} x_{27} x_{30} + k_{25} x_{31} \\
    & \dot{x}_{31} = k_{23} x_{29} x_{30} + k_{24} x_{27} x_{30} - k_{25} x_{31} \\
    & \dot{x}_{32} = -k_{26} x_7 x_{32} + k_{27} x_{33} \\
    & \dot{x}_{33} = k_{26} x_7 x_{32} - k_{27} x_{33} \\
    & \dot{x}_{34} = -k_{28} x_{29} x_{34} - k_{29} x_{28} x_{34} + k_{30} x_{35} \\
    & \dot{x}_{35} = k_{28} x_{29} x_{34} + k_{29} x_{28} x_{34} - k_{30} x_{35} \\
    & \dot{x}_{20} = -k_{31} x_{20} x_{35} + k_{32} x_{21} \\
    & \dot{x}_{21} = k_{31} x_{20} x_{35} - k_{32} x_{21} \\
    & \dot{x}_{36} = -k_{33} x_{31} x_{36} + k_{34} x_{37} \\
    & \dot{x}_{37} = k_{33} x_{31} x_{36} - k_{34} x_{37} \\
    & \dot{x}_{38} = -k_{35} x_{37} x_{38} - k_{36} x_{38} x_{41} + k_{37} x_{39} \\
    & \dot{x}_{39} = k_{35} x_{37} x_{38} + k_{36} x_{38} x_{41} - k_{37} x_{39} \\
    & \dot{x}_{40} = -k_{38} x_7 x_{40} - k_{39} x_{22} x_{40} - k_{40} x_{40} + k_{42} x_{42} \\
    & \dot{x}_{41} = k_{38} x_7 x_{40} + k_{39} x_{22} x_{40} + k_{40} x_{40} - k_{41} x_{41} \\
    & \dot{x}_{42} = k_{41} x_{41} - k_{42} x_{42} \\
    & \dot{x}_{43} = -k_{43} x_{41} x_{43} + k_{44} x_{44} \\
    & \dot{x}_{44} = k_{43} x_{41} x_{43} - k_{44} x_{44}
\end{align*}
where the following are concentrations:
\begin{align*}
    & x_2 = \text{Inactive insulin receptors} \\
    & x_3 = \text{Insulin-bound receptors} \\
    & x_4 = \text{Tyrosine-phosphorylated receptors} \\
    & x_6 = \text{Internalized dephosphorylated receptors} \\
    & x_7 = \text{Tyrosine-phosphorylated and internalized receptors} \\
    & x_9 = \text{Inactive IRS-1} \\
    & x_{10} = \text{Tyrosine-phosphorylated IRS-1} \\
    & x_{20} = \text{Intracellular GLUT4} \\
    & x_{21} = \text{Cell surface GLUT4} \\
    & x_{22} = \text{Combined tyrosine/serine 307-phosphorylated IRS-1} \\
    & x_{23} = \text{Serine 307-phosphorylated IRS-1} \\
    & x_{24} = \text{Inactive negative feedback} \\
    & x_{25} = \text{Active negative feedback} \\
    & x_{26} = \text{Inactive PKB} \\
    & x_{27} = \text{Threonine 308-phosphorylated PKB} \\
    & x_{28} = \text{Serine 473-phosphorylated PKB} \\
    & x_{29} = \text{Combined threonine 308/serine 473-phosphorylated PKB} \\
    & x_{30} = \text{mTORC1} \\
    & x_{31} = \text{mTORC1 involved in phosphorylation of IRS-1 at serine 307} \\
    & x_{32} = \text{mTORC2} \\
    & x_{33} = \text{mTORC2 involved in phosphorylation of PKB at threonine 473} \\
    & x_{34} = \text{AS160} \\
    & x_{35} = \text{AS160 phosphorylated at threonine 642} \\
    & x_{36} = \text{S6K} \\
    & x_{37} = \text{Activated S6K phosphorylated at threonine 389} \\
    & x_{38} = \text{S6} \\
    & x_{39} = \text{Activated S6 phosphorylated at serine 235 and serine 236} \\
    & x_{40} = \text{ERK} \\
    & x_{41} = \text{ERK phosphorylated at threonine 202 and tyrosine 204} \\
    & x_{42} = \text{ERK sequestered in an inactive pool} \\
    & x_{43} = \text{Elk1} \\
    & x_{44} = \text{Elk1 phosphorylated at serine 383}.
\end{align*}
Note that all the species concentrations are measured in \%. The total concentration of each of the following sets of species is 100\% at any given time: $\{ x_2, x_3, x_4, x_6, x_7 \}$, $\{ x_9, x_{10}, x_{22}, x_{23} \}$, $\{ x_{20}, x_{21} \}$, $\{ x_{24}, x_{25} \}$, $\{ x_{26}, \ldots, x_{29} \}$, $\{ x_{30}, x_{31} \}$, $\{ x_{32}, x_{33} \}$, $\{ x_{34}, x_{35} \}$, $\{ x_{36}, x_{37} \}$, $\{ x_{38}, x_{39} \}$, \linebreak $\{ x_{40}, x_{41}, x_{42} \}$, and $\{ x_{43}, x_{44} \}$.

We observe that the system has mass action kinetics since the exponent of every species concentration in the terms of the ODEs correspond to their stoichiometry in their respective reaction.

\section{Summary of the Properties of \\ INSMS and INRES}
\label{app:properties}

\setcounter{table}{0}
\renewcommand{\thetable}{C.\arabic{table}}

We provide in this section a table showing the different properties discussed in this paper.

Table \ref{tab:properties} summarizes the properties of INSMS and INRES.

\begin{table}[ht]
    \begin{center}
        \caption{Summary of the properties of INSMS and INRES\\}
        \label{tab:properties}
        \begin{tabular}{@{}|c|l|l|@{}}
            \hline
            \textbf{Property Class} & \multicolumn{1}{c|}{\textbf{INSMS}} & \multicolumn{1}{c|}{\textbf{INRES}} \\
            \hline
            \multirow{14}{*}{\shortstack{Network}} & \multicolumn{2}{c|}{Branching} \\
             \cline{2-3}
             & \multicolumn{2}{c|}{Closed} \\
             \cline{2-3}
             & \multicolumn{2}{c|}{High reactant diversity} \\
             \cline{2-3}
             & \multicolumn{2}{c|}{Non-cycle terminal} \\
             \cline{2-3}
             & \multicolumn{2}{c|}{Non-point terminal} \\
             \cline{2-3}
             & \multicolumn{2}{c|}{Not (weakly) reversible} \\
             \cline{2-3}
             & \multicolumn{2}{c|}{Positive dependent} \\
             \cline{2-3}
             & \multicolumn{2}{c|}{$t$-minimal} \\
             \cline{2-3}
             & 10 subnetworks & 12 subnetworks \\
             & Nonconservative & Conservative \\
             & Concordant & Discordant \\
             & 5 subnetworks with $\delta = 0$ & 0 subnetworks with $\delta = 0$ \\
             & 9 subnetworks with $s = 1$ & 8 subnetworks with $s = 1$ \\
             & 0 subnetworks with $\delta_{\rho} = 0$ & 12 subnetworks with $\delta_{\rho} = 0$ \\
            \hline
            \multirow{6}{*}{\shortstack{Structo-Kinetic}} & \multicolumn{2}{c|}{No complex balanced equilibria} \\
             \cline{2-3}
             & \multicolumn{2}{c|}{Kinetic and stoichiometric subspaces coincide} \\
             \cline{2-3}
             & \multicolumn{2}{c|}{Monostationary} \\
             \cline{2-3}
             & Injective & Non-injective \\
             & Monostationary in all weakly & Multistationary for some weakly \\
             & \hspace{0.25 cm} monotonic systems & \hspace{0.25 cm} monotonic systems \\
            \hline
            \multirow{2}{*}{Kinetic} & \multicolumn{2}{c|}{Nondegenerate equilibria} \\
             \cline{2-3}
             & 8 ACR species & No ACR species \\
            \hline
        \end{tabular}
    \end{center}
\end{table}

\end{appendices}


\begin{thebibliography}{99}

\bibitem{AJLM2017}
Arceo C, Jose E, Lao A, Mendoza E (2017) Reaction networks and kinetics of biochemical systems. Math Biosci 283:13--29. \begingroup\color{blue} \href{https://doi.org/10.1016/j.mbs.2016.10.004}{https://doi.org/10.1016/j.mbs.2016.10.004} \endgroup

\bibitem{BC2015}
Braatz, E, Coleman, R (2015) A mathematical model of insulin resistance in Parkinson's disease. Comput Biol Chem 56:84--97. \begingroup\color{blue} \href{https://doi.org/10.1016/j.compbiolchem.2015.04.003}{https://doi.org/10.1016/j.compbiolchem.2015.04.003} \endgroup

\bibitem{BNFBECS2013}
Br{\"a}nnmark C, Nyman E, Fagerholm S, Bergenholm L, Ekstrand E, Cedersund G, Str{\aa}lfors P (2013) Insulin signaling in type 2 diabetes: experimental and modeling analyses reveal mechanisms of insulin resistance in human adipocytes. J Biol Chem 288(14):9867--9880. \\ \begingroup\color{blue} \href{https://doi.org/10.1074/jbc.m112.432062}{https://doi.org/10.1074/jbc.m112.432062} \endgroup

\bibitem{CBHB2009}
Chellaboina V, Bhat S, Haddad W, Bernstein D (2009) Modeling and analysis of mass-action kinetics. IEEE Control Syst 29(4):60--78. \begingroup\color{blue} \href{https://doi.org/10.1109/MCS.2009.932926}{https://doi.org/10.1109/MCS.2009.932926} \endgroup

\bibitem{FML2021}
Fari\~{n}as H, Mendoza E, Lao A (2021) Chemical reaction network decompositions and realizations of S-systems. Philipp Sci Lett 14(1):147--157

\bibitem{FEIN1987}
Feinberg M (1987) Chemical reaction network structure and the stability of complex isothermal reactors: I. The deficiency zero and deficiency one theorems. Chem Eng Sci 42(10):2229--2268. \begingroup\color{blue} \href{https://doi.org/10.1016/0009-2509(87)80099-4}{https://doi.org/10.1016/0009-2509(87)80099-4} \endgroup

\bibitem{FEIN2019}
Feinberg M (2019) Foundations of chemical reaction network theory. Springer, Switzerland, \\ \begingroup\color{blue} \href{https://doi.org/10.1007/978-3-030-03858-8}{https://doi.org/10.1007/978-3-030-03858-8} \endgroup

\bibitem{CRNToolbox}
Feinberg M, Ellison P, Ji H, Knight D (2018) The Chemical Reaction Network Toolbox Version 2.35. \begingroup\color{blue} \href{https://doi.org/10.5281/zenodo.5149266}{https://doi.org/10.5281/zenodo.5149266} \endgroup

\bibitem{FEHO1977}
Feinberg M, Horn J (1977) Chemical mechanism structure and the coincidence of the stoichiometric and kinetic subspaces. Arch Ration Mech Anal 66(1):83--97. \\ \begingroup\color{blue} \href{https://doi.org/10.1007/bf00250853}{https://doi.org/10.1007/bf00250853} \endgroup

\bibitem{FOMF2021}
Fontanil L, Mendoza E, Fortun N (2021) A computational approach to concentration robustness in power law kinetic systems of Shinar-Feinberg type. MATCH Commun Math Comput Chem 86(3):489--516

\bibitem{FOME2023}
Fortun N, Mendoza E (2023) Comparative analysis of carbon cycle models via kinetic representations. J Math Chem 61:896--932. \begingroup\color{blue} \href{https://doi.org/10.1007/s10910-022-01442-8}{https://doi.org/10.1007/s10910-022-01442-8} \endgroup

\bibitem{HEDC2021}
Hernandez B, De la Cruz R (2021) Independent decompositions of chemical reaction networks. Bull Math Biol 83(76):1--23. \begingroup\color{blue} \href{https://doi.org/10.1007/s11538-021-00906-3}{https://doi.org/10.1007/s11538-021-00906-3} \endgroup

\bibitem{HLJK2022}
Hernandez B, Lubenia P, Johnston M, Kim J (2023) A framework for deriving analytic steady states of biochemical reaction networks. PLoS Comput Biol 19(4):e1011039. \begingroup\color{blue} \href{https://doi.org/10.1371/journal.pcbi.1011039}{https://doi.org/10.1371/journal.pcbi.1011039} \endgroup

\bibitem{HEME2021}
Hernandez B, Mendoza E (2021) Positive equilibria of Hill-type kinetic systems. J Math Chem 59:840--870. \begingroup\color{blue} \href{https://doi.org/10.1007/s10910-021-01230-w}{https://doi.org/10.1007/s10910-021-01230-w} \endgroup

\bibitem{HWDLC2014}
Huang C, Wu M, Du J, Liu D, Chan C (2014) Systematic modeling for the insulin signaling network mediated by IRS1 and IRS2. J Theor Biol 355:40--52. \begingroup\color{blue} \href{https://doi.org/10.1016/j.jtbi.2014.03.030}{https://doi.org/10.1016/j.jtbi.2014.03.030} \endgroup

\bibitem{JB2019}
Johnston M, Burton E (2019) Computing weakly reversible deficiency zero network translations using elementary flux modes. Bull Math Biol 81:1613--1644. \begingroup\color{blue} \href{https://doi.org/10.1007/s11538-019-00579-z}{https://doi.org/10.1007/s11538-019-00579-z} \endgroup

\bibitem{JMP2019}
Johnston M, M\"{u}ller S, Pantea C (2019) A deficiency-based approach to parameterizing positive equilibria of biochemical reaction systems. Bull Math Biol 81:1143--1172. \\ \begingroup\color{blue} \href{https://doi.org/10.1007/s11538-018-00562-0}{https://doi.org/10.1007/s11538-018-00562-0} \endgroup

\bibitem{KPDDG2012}
Karp R, P\'{e}rez-Mill\'{a}n M, Dasgupta T, Dickenstein A, Gunawardena J (2012) Complex-linear invariants of biochemical networks. J Theor Biol 311:130--138. \\ \begingroup\color{blue} \href{https://doi.org/10.1016/j.jtbi.2012.07.004}{https://doi.org/10.1016/j.jtbi.2012.07.004} \endgroup

\bibitem{LML2022}
Lubenia P, Mendoza E, Lao A (2022) Reaction network analysis of metabolic insulin signaling. Bull Math Biol 84(129):1--22. \begingroup\color{blue} \href{https://doi.org/10.1007/s11538-022-01087-3}{https://doi.org/10.1007/s11538-022-01087-3} \endgroup

\bibitem{MRBH2015}
MacLean A, Rosen Z, Byrne H, Harrington H (2015) Parameter-free methods distinguish Wnt pathway models and guide design of experiments. Proc Natl Acad Sci USA 112(9):2652--2657. \begingroup\color{blue} \href{https://doi.org/10.1073/pnas.1416655112}{https://doi.org/10.1073/pnas.1416655112} \endgroup

\bibitem{MEST2022}
Meshkat N, Shiu A, Torres A (2022) Absolute concentration robustness in networks with low-dimensional stoichiometric subspace. Vietnam J Math \\  \begingroup\color{blue} \href{https://doi.org/10.1007/s10013-021-00524-5}{https://doi.org/10.1007/s10013-021-00524-5} \endgroup

\bibitem{NTNLV2020}
Nguyen T, Ta Q, Nguyen T, Le T, Vo V (2020) Role of insulin resistance in the Alzheimer's disease progression. Neurochem Res 45:1481--1491. \begingroup\color{blue} \href{https://doi.org/10.1007/s11064-020-03031-0}{https://doi.org/10.1007/s11064-020-03031-0} \endgroup

\bibitem{NRFBCS2014}
Nyman E, Rajan M, Fagerholm S, Br\"{a}nnmark C, Cedersund G, Str{\aa}lfors P (2014) A single mechanism can explain network-wide insulin resistance in adipocytes from obese patients with type 2 diabetes. J Biol Chem 289(48):33215--33230. \begingroup\color{blue} \href{https://doi.org/10.1074/jbc.M114.608927}{https://doi.org/10.1074/jbc.M114.608927} \endgroup

\bibitem{ONEASZ2018}
Ormazabal V, Nair S, Elfeky O, Aguayo C, Salomon C, Zu{\~n}iga F (2018) Association between insulin resistance and the development of cardiovascular disease. Cardiovasc Diabetol 17:1--14. \begingroup\color{blue} \href{https://doi.org/10.1186/s12933-018-0762-4}{https://doi.org/10.1186/s12933-018-0762-4} \endgroup

\bibitem{QC1991}
Quon, Michael J. and Campfield, L. Arthur (1991) A mathematical model and computer simulation study of insulin receptor regulation. J Theor Biol 150(1):59--72. \begingroup\color{blue} \href{https://doi.org/10.1016/S0022-5193(05)80475-8}{https://doi.org/10.1016/S0022-5193(05)80475-8} \endgroup

\bibitem{SSQ2002}
Sedaghat A, Sherman A, Quon M (2002) A mathematical model of metabolic insulin signaling pathways. Am J Physiol Endocrinol Metab 283(5):E1084--E1101. \\ \begingroup\color{blue} \href{https://doi.org/10.1152/ajpendo.00571.2001}{https://doi.org/10.1152/ajpendo.00571.2001} \endgroup

\bibitem{SHFE2010}
Shinar G, Feinberg M (2010) Structural sources of robustness in biochemical reaction networks. Science 327(5971):1389--1391. \begingroup\color{blue} \href{https://doi.org/10.1126/science.1183372}{https://doi.org/10.1126/science.1183372} \endgroup

\bibitem{SHFE2011}
Shinar G, Feinberg M (2011) Design principles for robust biochemical reaction networks: What works, what cannot work, and what might almost work. Math Biosci 231(1):39--48. \\ \begingroup\color{blue} \href{https://doi.org/10.1016/j.mbs.2011.02.012}{https://doi.org/10.1016/j.mbs.2011.02.012} \endgroup

\bibitem{SHFE2012}
Shinar G, Feinberg M (2012) Concordant chemical reaction networks. Math Biosci 240(2):92--113. \begingroup\color{blue} \href{https://doi.org/10.1016/j.mbs.2012.05.004}{https://doi.org/10.1016/j.mbs.2012.05.004} \endgroup

\bibitem{SHFE2013}
Shinar G, Feinberg M (2013) Concordant chemical reaction networks and the species-reaction graph. Math Biosci 241(1):1--23. \begingroup\color{blue} \href{https://doi.org/10.1016/j.mbs.2012.08.002}{https://doi.org/10.1016/j.mbs.2012.08.002} \endgroup

\bibitem{TGCCLMOF2020}
Tanase D, Gosav E, Costea C, Ciocoiu M, Lacatusu C, Maranduca M, Ouatu A, Floria M (2020) The intricate relationship between type 2 diabetes mellitus (T2DM), insulin resistance (IR), and nonalcoholic fatty liver disease (NAFLD). J Diabetes Res 2020:1--16. \begingroup\color{blue} \href{https://doi.org/10.1155/2020/3920196}{https://doi.org/10.1155/2020/3920196} \endgroup

\bibitem{TG2009a}
Thomson M, Gunawardena J (2009) The rational parameterisation theorem for multisite post-translational modification systems. J Theor Biol 261(4):626--636. \begingroup\color{blue} \href{https://doi.org/10.1016/j.jtbi.2009.09.003}{https://doi.org/10.1016/j.jtbi.2009.09.003} \endgroup

\bibitem{TG2009b}
Thomson M, Gunawardena J (2009) Unlimited multistability in multisite phosphorylation systems. Nature 460, 274--277. \begingroup\color{blue} \href{https://doi.org/10.1038/nature08102}{https://doi.org/10.1038/nature08102} \endgroup

\bibitem{VLMA2019}
Villar J, Lubenia P, Mendoza E, Arceo C (2019) Structural stability analysis of models of dopamine synthesis and D1 receptor trafficking in RPT cells using CRNT. Philipp J Sci 148(3):523–533

\bibitem{WQ2000}
Wanant S, Quon M (2000) Insulin receptor binding kinetics: Modeling and simulation studies. J Theor Biol 205(3):355--364. \begingroup\color{blue} \href{https://doi.org/10.1006/jtbi.2000.2069}{https://doi.org/10.1006/jtbi.2000.2069} \endgroup

\bibitem{YFBS2019}
Yaribeygi H, Farrokhi F, Butler A, Sahebkar A (2019) Insulin resistance: Review of the underlying molecular mechanisms. J Cell Physiol 234(6):8152--8161. \begingroup\color{blue} \href{https://doi.org/10.1002/jcp.27603}{https://doi.org/10.1002/jcp.27603} \endgroup

\end{thebibliography}
\end{document}